\documentclass[12pt]{article}
\usepackage{amsfonts}
\usepackage{color}
\usepackage{graphicx}
\usepackage{amsmath}
\usepackage{float}
\usepackage{amssymb}
\providecommand{\U}[1]{\protect\rule{.1in}{.1in}}
\textwidth=6.5in \hoffset=-.75in \voffset=-1in \numberwithin{equation}{section}
\numberwithin{figure}{section}

\newcommand {\be}{\begin{equation}}
 \newcommand {\ee}{\end{equation}}
 \newcommand {\bea}{\begin{array}}
 
 \newcommand {\eea}{\end{array}}

\begin{document}

\begin{titlepage}
\bigskip \begin{flushright}
\end{flushright}
\vspace{1cm}
\begin{center}
{\Large \bf {Vector Fields in Kerr/CFT Correspondence}}\\
\vskip 1cm
\end{center}
\vspace{1cm}
\begin{center}
A. M.
Ghezelbash{ \footnote{amg142@campus.usask.ca}}, H. M. Siahaan{\footnote{hms923@campus.usask.ca}}
\\
Department of Physics and Engineering Physics, \\ University of Saskatchewan,
Saskatoon, Saskatchewan S7N 5E2, Canada\\
\vspace{1cm}
\end{center}
\begin{abstract}

In this paper, we use the appropriate boundary action for the vector fields near the horizon of near extremal Kerr black hole to calculate the two-point function for the vector fields in Kerr/CFT correspondence.  We show the gauge-independent part of the two-point function is in agreement with what is expected from CFT.

\end{abstract}
\bigskip
\end{titlepage}\onecolumn

\bigskip

\section{Introduction}
\par

A decade after the proposal of AdS/CFT, another type of holography namely Kerr/CFT correspondence appeared \cite{Guica:2008mu}. The Kerr/CFT correspondence provides us a new type of holography that may have a connection to the realistic astrophysical objects. In the early days of Kerr/CFT correspondence (see for example \cite{extremeKerr-CFT}) the discussions were concentrated to the extremal or near extremal case, which are still realistic since the X-ray astrophysical observations support the existence of near extremal rotating black hole candidates \cite{spin-GRS}. The $SL(2,\mathbb{R})\times U(1)$ symmetry of near horizon extreme Kerr (NHEK) hints the existence of a dual CFT. The dual CFT can describe some of physical aspects of black holes in near horizon. The authors \cite{Guica:2008mu} realized this idea and employed the asymptotic symmetry group (ASG) technique to show the corresponding conformal symmetry and obtain the central charges of dual CFT. The corresponding central charges as well as the Frolov-Thorne temperatures of CFT, give the exact microscopic entropy by using CFT Cardy formula. The microscopic CFT entropy is in perfect agreement with the Bekenstein-Hawking entropy of extremal Kerr black hole that indeed establishes the Kerr/CFT correspondence. 
\par
A more general proposal for Kerr/CFT correspondence, which is not limited to extremal rotating black holes, was appeared in \cite{castro}. The proposal has been applied to several types of generic rotating black holes \cite{hiddenkerrcft}. In generic Kerr/CFT correspondence, the conformal symmetry for the generic rotating black holes (also known as hidden conformal symmetry) is revealed from the solution space of the wave function of scalar (or higher spin) fields. It is unlike the extremal Kerr/CFT correspondence where the conformal symmetry arises from the spacetime structure. In both types of extremal and generic Kerr/CFT correspondence, the results for some physical quantities of rotating black holes, such as the Bekenstein-Hawking entropy and scattering cross section, are in perfect agreement with the corresponding physical quantities in the dual CFT.
\par
Almost for all generic rotating black holes, the hidden conformal symmetry has been found by looking at the symmetry of solution space of a scalar test particle. However higher spin test particles in Kerr/CFT correspondence also were considered recently in \cite{Hartman:2009nz,Chen:2010ni}, tough the higher spin test particles in the background of black holes had considered before \cite{p1,p2} with different techniques.
Quite recently the authors of \cite{Becker:2012vda} found the two-point function of spinor fields in Kerr/CFT correspondence by variation of boundary action for spin-1/2 particles. They determined an appropriate boundary term for the spinors in NHEK geometry and used it to calculate the two-point function of spinors. Moreover they found a relation between spinors in the four-dimensional bulk and the boundary spinors living in two dimensions. 
The two-point function of spinor fields is in agreement with the correlation function of a two-dimensional CFT. The variational method in \cite{Becker:2012vda}  for spinors in NHEK geometry is in the same spirit for  spinors in the context of AdS/CFT correspondence \cite{mueck1}.
In reference \cite{Hartman:2009nz}, the authors show that the two-point function of an operator at left and right temperatures $(T_L,T_R)$ and with conformal dimensions $(h_L,h_R)$, is
\be
G \sim \left( { - 1} \right)^{h_L  + h_R } \left( {\frac{{\pi T_L }}
{{\sinh \left( {\pi T_L t^ -  } \right)}}} \right)^{2h_L } \left( {\frac{{\pi T_R }}
{{\sinh \left( {\pi T_R t^ +  } \right)}}} \right)^{2h_R } \label{GfromConf}
\ee
that will be useful later in this article. 
\par 
In this article, inspired by Becker et al \cite{Becker:2012vda}, 
we derive the two-point function for Maxwell fields in Kerr spacetime by varying the corresponding boundary action. Unlike the analysis of spin-1/2 particles that there is no gauge condition, one needs to perform a more careful treatment for the gauge fields where they are subjected to the gauge condition. In this regard, we use the wave equation for spin-1 objects in Kerr background given in \cite{Teukolsky:1973ha}. We note that in \cite{Teukolsky:1973ha}, Teukolsky derived a set of wave equations for spin-$0, 1/2, 1$, and $2$ field perturbations in Kerr background. Moreover, Chandrasekhar derived the solutions for Maxwell fields in Kerr spacetime in term of Teukolsky radial and angular wave functions \cite{Chandrasekhar1976mk}. The gauge condition that is used in \cite{Chandrasekhar1976mk} to get the spin-1 field solutions in Kerr background is quite complicated which encumbers to derive the two-dimensional CFT correlators of vector fields.
However, we can use (\ref{GfromConf}) to justify that gauge-independent part of two-point function for Maxwell fields in NHEK geometry is dual to the thermal CFT correlators. 
\par
We start with the Maxwell action in four-dimensional Kerr background where all four components of Maxwell fields  are alive. After explicitly calculating the appropriate boundary action for the Maxwell fields, the leading terms in the boundary action contain only the boundary fields corresponding to $A_t$ and $A_\phi$.
 Interestingly enough, this result provides the correct number of degrees of freedom for the boundary fields and yields the corresponding two-point function of spin-1 fields. Both the dimensionless Hawking temperature $\tau_H$ and the boundary value of metric function $\Delta = (r-r_+)(r-r_-)$ are small numbers that play important role to get the appropriate number of Maxwell fields on the boundary. The smallness of these quantities is a result of considering the near horizon near extremal limit of Kerr geometry.
All the results of this article support the Kerr/CFT correspondence where four-dimensional rotating black holes physics is dual to two-dimensional CFT on the boundary.
\par
The organization of this paper is as follows. In section \ref{sec-Chandra}, we review briefly the solutions for the Maxwell fields in Kerr spacetime.
In section \ref{sec-bound}, we derive the proper boundary action for the Maxwell fields in Kerr spacetime and present it in a matrix form to facilitate the explicit calculation of the boundary action in next sections. 
In section \ref{sec-app},  we employ some approximations to simplify our analysis of the boundary action, namely we set the radial coordinate $r$ in some parts of the Maxwell fields to be equal to the boundary radius in ``near horizon'' region. 
The Teukolsky radial function $R(r)$ then describes the dependence of Maxwell field solutions to the radial coordinate. 
Then in section \ref{sec-2ptvarMax}, we derive the two-point function for Maxwell fields on the boundary surface by varying the corresponding boundary action. In section \ref{sec-CFT}, we find that the gauge-independent part of the two-point function can be obtained properly from the dual CFT. In section \ref{sec-conclu}, we present the conclusions and the other open questions.

\section{Spin-1 fields in the background of Kerr black holes}\label{sec-Chandra}
\subsection{Construction of solutions in Newman-Penrose formalism}

In this section, we briefly review the derivation of solutions to Maxwell equations in the background of Kerr black hole
\cite{Chandrasekhar:1985kt} and fix the notation in the article 
\footnote{There is a slight difference on some notations in constructing the solutions to Maxwell's equations in the background of Kerr black hole in literature such as
\cite{Chandrasekhar1976mk} and \cite{Chandrasekhar:1985kt}. In this paper, we mainly follow \cite{Chandrasekhar:1985kt}.}. In Boyer-Lindquist coordinate, the Kerr metric read as
\be
ds^2  =  - \frac{\Delta }
{{\rho ^2 }}\left( {dt - a\sin ^2 \theta d\phi } \right)^2  + \frac{{\rho ^2 }}
{\Delta }dr^2  + \rho ^2 d\theta ^2  + \frac{{\sin ^2 \theta }}
{{\rho ^2 }}\left( {adt - \left( {r^2  + a^2 } \right)d\phi } \right)^2  \label{Kerr-BLmetric},
\ee
where $\rho^2  = r^2  + a^2 \cos ^2 \theta $ and $\Delta  = r^2  + a^2  - 2Mr$. For later convenience, the corresponding contravariant components of the metric tensor for (\ref{Kerr-BLmetric}) are given by
\begin{equation}
  g^{rr}  = \frac{\Delta }
{{\rho ^2 }},~~g^{\theta \theta }  = \frac{1}
{{\rho ^2 }},~~g^{tt}  = \frac{{\left( {\Delta a^2 \sin ^2 \theta  - \left( {r^2  + a^2 } \right)^2 } \right)}}
{{\Delta \rho ^2 }}, \label{contra-metric1}\end{equation}\begin{equation} 
  g^{t\phi }  = \frac{{ - 2Mra}}
{{\Delta \rho ^2 }},~~g^{\phi \phi }  = \frac{{\left( {\Delta  - a^2 \sin ^2 \theta } \right)}}
{{\Delta \rho ^2 \sin ^2 \theta }}. \label{contra-metric2}
\end{equation}
Stationary and axisymmetric properties of the Kerr black hole suggest 
that the solution to Maxwell equations in this spacetime can be written as a superposition of waves with different frequencies $\omega$ and different periods $2m\pi$, $m = 0,1,2,...$ for coordinate $\phi$. Thus, the existence of Killing vectors $\partial_t$ and $\partial_\phi$ 
for Kerr spacetime (\ref{Kerr-BLmetric}) enable us to write down the dependence of spin-1 field solutions to $t$ and $\phi$ coordinates as $e^{ - i\omega t + im\phi }$.
\par
In his seminal work \cite{Teukolsky:1973ha}, Teukolsky showed that the equations of motions for the fields (with different spin weights) in Kerr background 
are separable in radial and angular directions.
In Newman-Penrose (NP) formalism, the real null-vectors $l^\mu$ and $n^\mu$ and the complex null-vector $m^\mu$ for Kerr spacetime (\ref{Kerr-BLmetric}) are given by \cite{Chandrasekhar:1985kt}
\begin{equation}
l^\mu = \Delta ^{ - 1} \left( {r^2  + a^2 ,\Delta ,0,a} \right),\label{ll}
\end{equation}
\begin{equation}
n^\mu = \frac{1}{{2\rho ^2 }}\left( {r^2  + a^2 , - \Delta ,0,a} \right),\label{nn}
\end{equation}
\begin{equation}
m^\mu = \frac{1}{{\bar \rho \sqrt 2 }}\left( {ia\sin \theta ,0,1,\frac{i}{{\sin \theta }}} \right),\label{mm}
\end{equation}
in $(t,r,\theta,\phi)$ coordinate system where
$\bar \rho  = r + ia\cos \theta$ and $\bar \rho * = r - ia\cos \theta$.
Contracting the vectors $l^\mu,n^\mu$ and $m^\mu$ by $\partial_\mu$, we get the following differential operators 
\begin{equation}
l = \mathcal{D}_0 ~~,~~
n =  - \frac{\Delta }{{2\rho ^2 }}\mathcal{D}^\dag  _0,
\label{ln}
\end{equation}
\begin{equation}
m = \frac{1}{{\bar \rho \sqrt 2 }}\mathcal{L}^\dag  _0.
\label{mmbar}
\end{equation}
We also  consider the operator
\begin{equation}
\bar m = \frac{1}{{\bar \rho ^* \sqrt 2 }}\mathcal{L}_0.
\label{mbarop}
\end{equation}
The differential operators (\ref{ln}), (\ref{mmbar}) and (\ref{mbarop}) act on any function that its dependence on coordinates $t$ and $\phi$ is given by $e^{ - i\omega t + im\phi }$. The operators $\mathcal{D} _0$, $\mathcal{D}^\dag  _0$, $\mathcal{L}  _0 $ and $\mathcal{L}^\dag  _0 $ are special cases of 
\begin{equation}
\mathcal{D}_n  = \frac{\partial }{{\partial r}} + \frac{{iK}}{\Delta } + 2n\left( {\frac{{r - M}}{\Delta }} \right) ~~~,~~~ \mathcal{D}_n^\dag  = \frac{\partial }{{\partial r}} - \frac{{iK}}{\Delta } + 2n\left( {\frac{{r - M}}{\Delta }} \right),
\label{DDdagger}
\end{equation}
\begin{equation}
\mathcal{L}_n  = \frac{\partial }{{\partial \theta }} + Q + n\cot \theta ~~~,~~~ \mathcal{L}_n^\dag  = \frac{\partial }{{\partial \theta }} - Q + n\cot \theta,
\label{LLdagger}
\end{equation}
where  $K$ and $Q$ are given by 
\be
K =  - \left( {r^2  + a^2 } \right)\omega  + am,\label{Kdef}
\ee 
and 
\be 
Q =  - a\omega \sin \theta  + m\left( {\sin \theta } \right)^{ - 1}\label{Qdef},
\ee
and $n \in \mathbb{Z}$. As we notice, the operators $\mathcal{D}_n$  and $\mathcal{D}_n^\dag$ are purely radial dependent operators, whereas $\mathcal{L}_n$ and $\mathcal{L}_n^\dag$ are purely angular dependent operators.
\par
Contracting the field-strength tensor $F_{\mu\nu}$ with the basis vectors (\ref{ll}) - (\ref{mm}) yield three complex scalars $\Phi_0$, $\Phi_1$ and $\Phi_2$ which can be read as
\begin{align}
\Phi _0  = F_{\mu \nu } l^\mu  m^\nu  \label{Phi0F},
\end{align}
\begin{align}
\Phi _1  = \frac{{\bar \rho ^* }}{{\sqrt 2 }}F_{\mu \nu } \left( {l^\mu  n^\nu   + \bar m^\mu  m^\nu  } \right) \label{Phi1F},
\end{align} and
\begin{align}
\Phi _2  = 2\left( {\bar \rho ^* } \right)^2 F_{\mu \nu } \bar m^\mu  n^\nu \label{Phi2F}. 
\end{align} 
The Maxwell's equations in the background (\ref{Kerr-BLmetric}) are given by
\begin{align}
g^{\mu \nu } \nabla _\nu  F_{\mu \rho }  = 0 \label{eom2},
\end{align} 
along with the Bianchi identity 
\begin{align}
\nabla _\mu  F_{\nu \rho }  + \nabla _\rho  F_{\mu \nu }  + \nabla _\nu  F_{\rho \mu }  = 0. \label{eom1}
\end{align}
Equation (\ref{eom1}) indicates that there is no source for Maxwell fields in the gravitational background (\ref{Kerr-BLmetric}).
Inserting all the spin coefficients and directional derivatives into Maxwell's equations gives a set of four equations 
in NP formalism
\begin{align}
\left( {\mathcal{L}_1  - \frac{{ia\sin \theta }}{{\bar \rho ^ *  }}} \right)\Phi _0  = \left( {\mathcal{D}_0  + \frac{1}{{\bar \rho ^ *  }}} \right)\Phi _1 \label{eq1},
\end{align}
\begin{align}
\left( {\mathcal{L}_0  + \frac{{ia\sin \theta }}{{\bar \rho ^ *  }}} \right)\Phi _1  = \left( {\mathcal{D}_0  - \frac{1}{{\bar \rho ^ *  }}} \right)\Phi _2 \label{eq2},
\end{align}
\begin{align}
\left( {\mathcal{L}^\dag  _1  - \frac{{ia\sin \theta }}{{\bar \rho ^ *  }}} \right)\Phi _2  =  - \Delta \left( {\mathcal{D}^\dag  _0  + \frac{1}{{\bar \rho ^ *  }}} \right)\Phi _1 \label{eq3},
\end{align} and
\begin{align}
\left( {\mathcal{L}^\dag  _0  + \frac{{ia\sin \theta }}{{\bar \rho ^ *  }}} \right)\Phi _1  =  - \Delta \left( {\mathcal{D}^\dag  _1  - \frac{1}{{\bar \rho ^ *  }}} \right)\Phi _0 \label{eq4}.
\end{align}
The equations (\ref{eq1}) - (\ref{eq4}) can be decoupled to two differential equations for  $\Phi _0$ and $\Phi _2$ by noticing that two operators
\begin{eqnarray}
Y_m  &=& \mathcal{D} + m\left( {\bar \rho ^ *  } \right)^{ - 1} \label{Ymdef},\\
Z_m  &=& \mathcal{L} + ima\sin \theta \left( {\bar \rho ^ *  } \right)^{ - 1},\label{Zmdef}
\end{eqnarray}
commute, i.e. $\left[ {Y_m ,Z_n } \right] = 0$.
In (\ref{Ymdef}) and (\ref{Zmdef}), $\mathcal{D}$ can be either $\mathcal{D} _n$ or $\mathcal{D}^\dag _n$ and $\mathcal{L}$ can be either $\mathcal{L} _n$ or $\mathcal{L}^\dag _n$ respectively. The two decoupled differential equations for $\Phi_0$ and $\Phi_2$ are 
\begin{align}
	\left[ {\left( {\mathcal{L}^\dag  _0  + \frac{{ia\sin \theta }}{{\bar \rho ^ *  }}} \right)\left( {\mathcal{L}_1  - \frac{{ia\sin \theta }}{{\bar \rho ^ *  }}} \right) + \Delta \left( {\mathcal{D}_1  + \frac{1}{{\bar \rho ^ *  }}} \right)\left( {\mathcal{D}^\dag  _1  - \frac{1}{{\bar \rho ^ *  }}} \right)} \right]\Phi _0  = 0,\label{Phi0}
\end{align}
and
\begin{align}
	\left[ {\left( {\mathcal{L}_0  + \frac{{ia\sin \theta }}{{\bar \rho ^ *  }}} \right)\left( {\mathcal{L}^\dag  _1  - \frac{{ia\sin \theta }}{{\bar \rho ^ *  }}} \right) + \Delta \left( {\mathcal{D}_0  + \frac{1}{{\bar \rho ^ *  }}} \right)\left( {\mathcal{D}_0  - \frac{1}{{\bar \rho ^ *  }}} \right)} \right]\Phi _2  = 0.\label{Phi2}
\end{align}
We notice that to obtain equation (\ref{Phi0}), we have used the identity $\mathcal{D}_0 \Delta  = \Delta \mathcal{D}_1$. 
Using the identities (\ref{ID1}), (\ref{ID2}), (\ref{ID3}) and (\ref{ID4}) in appendix A, we can simplify equations (\ref{Phi0}) and (\ref{Phi2}) to
\begin{align}
	\left( {\Delta \mathcal{D}_1 \mathcal{D}^\dag  _1  + \mathcal{L}^\dag  _0 \mathcal{L}_1  + 2i\omega \left( {r + ia\cos \theta } \right)} \right)\Phi _0  = 0,\label{PPhi0}
\end{align}
and
\begin{align}
	\left( {\Delta \mathcal{D}^\dag  _0 \mathcal{D}_0  + \mathcal{L}_0 \mathcal{L}^\dag  _1  - 2i\omega \left( {r + ia\cos \theta } \right)} \right)\Phi _2  = 0.\label{PPhi2}
\end{align}
The equations (\ref{PPhi0}) and (\ref{PPhi2}) are clearly separable in 
$r$ and $\theta$ 
and called the Teukolsky equations for the massless particles with spin weight 1. For convenience, we set
\be
\Phi _0  = \Psi _ +,  ~~~~ \Phi _2  = \Psi _ -,\label{PhitoTeu}
\ee
where $\Psi_\pm \equiv S_ \pm \left( \theta  \right)R_ \pm(r)$ and $R_{ \pm} \left( r \right)$ and $S_{ \pm} \left( \theta \right)$ are functions of $r$ and $\theta$ only, respectively. The functions $\Psi_\pm$ contain the $r$ and $\theta$ dependence of Teukolsky wave functions $\tilde\Psi _ \pm$ for Maxwell field perturbation with spin weights $\pm 1$ 
\be
\tilde\Psi _ \pm   = e^{ - i\omega t + im\phi } R_ \pm  \left( r \right)S_ \pm \left( \theta  \right) \equiv e^{ - i\omega t + im\phi } \Psi_\pm.\label{PsiTeu}
\ee
Plugging (\ref{PhitoTeu}) into equations (\ref{PPhi0}) and (\ref{PPhi2}) we obtain a set of equations
\begin{align}
\left( {\Delta \mathcal{D}_1 \mathcal{D}^\dag  _1  + 2i\omega } \right)R_+  = \lambda R_+ \label{Teu1},
\end{align}
\begin{align}
\left( {\mathcal{L}^\dag  _0 \mathcal{L}_1  - 2a\omega \cos \theta } \right)S_+  =  - \lambda S_+,\label{Teu2}
\end{align}
and
\begin{align}
\left( {\Delta \mathcal{D}_0 \mathcal{D}^\dag  _0  - 2i\omega } \right)R_-  = \lambda R_-, \label{Teu3}
\end{align}
\begin{align}
\left( {\mathcal{L}_0 \mathcal{L}^\dag  _1  + 2a\omega \cos \theta } \right)S_-  =  - \lambda S_-, \label{Teu4}
\end{align}
for the radial $R_{\pm}$ and angular $S_{\pm}$ functions where $\lambda$ is the separation constant. 
The radial solutions to Teukolsky equations have been found in reference \cite{Mano:1996gn}. 
The radial solutions also have been obtained for near horizon near extremal Kerr in reference \cite{Hartman:2009nz} by taking near and far region limits of a generic Teukolsky equation \cite{Teukolsky:1973ha}. 
Applying the operator $\left( {\mathcal{L}_0  + ia\bar \rho ^{* - 1} \sin \theta } \right)$ to (\ref{eq1}) and $\left( {\mathcal{D}_0  + \bar \rho ^{* - 1} } \right)$ to (\ref{eq2}) and adding them up, we find
\begin{align}
\left( {\mathcal{L}_0  + \frac{{ia\sin \theta }}{{\bar \rho ^ *  }}} \right)\left( {\mathcal{L}_1  - \frac{{ia\sin \theta }}{{\bar \rho ^ *  }}} \right)\Phi _0  = \left( {\mathcal{D}_0  + \frac{1}{{\bar \rho ^ *  }}} \right)\left( {\mathcal{D}_0  - \frac{1}{{\bar \rho ^ *  }}} \right)\Phi _2. \label{P1P2}
\end{align}
Furthermore equation (\ref{P1P2}) simplifies to
\begin{align}
	\mathcal{L}_0 \mathcal{L}_1 \Phi _0  = \mathcal{D}_0 \mathcal{D}_0 \Phi _2 \label{P1P2s}.
\end{align}
As we notice from (\ref{PhitoTeu}) and (\ref{PsiTeu}), the complex scalars $\Phi_0$ and $\Phi_2$ are separable functions in terms of coordinates $r$ and $\theta$. Plugging the identifications (\ref{PhitoTeu}) into equation (\ref{P1P2s}), we get the equation
\begin{align}
	\frac{{\mathcal{L}_0 \mathcal{L}_1 S_+ }}{{S_- }} = \frac{{\Delta \mathcal{D}_0 \mathcal{D}_0 R_- }}{{\Delta R_+ }}\label{RSRS1},
\end{align}
that leads to
\begin{align}
	\mathcal{D}_0 \mathcal{D}_0 R_-  = CR_+ \label{RpRm},
\end{align}
which is one of the Teukolsky - Starobinsky identities \cite{Chandrasekhar:1985kt}. In (\ref{RpRm}), $C$ is the Starobinsky constant which in general cab be complex valued,
\begin{align}
\left| C \right|^2 = {\lambda^2  - 4\alpha ^2 \omega ^2 }\label{constC},	
\end{align}
and $\alpha$ is defined as
\be
\alpha^2 = a^2 - \frac{am}{\omega}\label{alpha}.
\ee
Moreover, for later convenience, we consider the angular functions ${S_+ }$ and ${S_- }$ normalized to unity 
\begin{align}
	\int\limits_0^\pi  {S_+^2 \sin \theta d\theta }  = \int\limits_0^\pi  {S_-^2 \sin \theta d\theta }  = 1.\label{normS}
\end{align}

\subsection{Chandrasekhar's solutions for Maxwell fields in Kerr background}
In this section, we derive in detail the solutions to Maxwell's equations in Kerr background by using the three complex scalars (\ref{Phi0F}), (\ref{Phi1F}) and (\ref{Phi2F}) that are related to Maxwell field strength tensor $F_{\mu \nu }  = \partial _\mu  A_\nu   - \partial _\nu  A_\mu $. We consider the gauge field $A_\mu$ as $\left( {A_t ,A_r ,A_\theta  ,A_\phi  } \right)$ in spherical coordinates. The complex scalars $\Phi_0$ and $\Phi_2$ given by (\ref{Phi0F}) and (\ref{Phi2F}), can be written as
\begin{eqnarray}
	\Phi _0  &=& \frac{1}{{\bar \rho   \sqrt 2 }}\left( {\mathcal{L}^\dag  _0 \left( {\frac{{r^2  + a^2 }}{\Delta }A_t  + A_r  + \frac{a}{\Delta }A_\phi  } \right) - \mathcal{D}_0 \left( {iaA_t \sin \theta  + A_\theta   + \frac{{iA_\phi  }}{{\sin \theta }}} \right)} \right),\label{Phi0A}\\
\Phi _2  &=& - \frac{{1 }}{{\bar \rho  \sqrt 2 }}\left( {\Delta \mathcal{D}^\dag  _0 \left( { - iaA_t \sin \theta  + A_\theta   - \frac{{iaA_\phi  }}{{\sin \theta }}} \right) + \mathcal{L}_0 \left( { - \Delta A_r  + \left( {r^2  + a^2 } \right)A_t  + aA_\phi  } \right)} \right).\nonumber \\
&&\label{Phi2A}
\end{eqnarray}
To simplify some expressions that will be handled hereafter, we define the following functions
\begin{align}
\Delta F_+  = \left( {r^2  + a^2 } \right)A_t  + \Delta A_r  + aA_\phi
~~,~~\Delta F_{ - }  = \left( {r^2  + a^2 } \right)A_t  - \Delta A_r  + aA_\phi  \label{FtoA}~,
\end{align}
\begin{align}
G_+  = iaA_t \sin \theta  + A_\theta   + i\frac{A_\phi}{\sin \theta}
~~,~~ G_-  =  - iaA_t \sin \theta  + A_\theta   - i\frac{A_\phi}{\sin \theta}  \label{GtoA}~.
\end{align}
Moreover, the following definitions would also be helpful \cite{Chandrasekhar:1985kt}
\begin{align}
	\xi_+ \left( r \right) = C^{ - 1} \left( {r\mathcal{D}_0  - 1} \right)R_-
~~~,~~~	\xi_- \left( r \right) = C^{ - 1} \left( {r\mathcal{D}^\dag  _0  - 1} \right)\left( {\Delta R_+ } \right) \label{xi},
\end{align}
\begin{align}
	\zeta_+ \left( \theta  \right) = C^{ - 1} \left( {\cos \theta \mathcal{L}^\dag  _1  + \sin \theta } \right)S_-
~~~,~~~	\zeta_- \left( \theta  \right) = C^{ - 1} \left( {\cos \theta \mathcal{L}_1  + \sin \theta } \right)S_+ \label{zeta},
\end{align}
where $C$ is the Starobinksy constant (\ref{constC}). 
\par
One can easily verify that the $r$-dependent functions $\xi _ \pm$ and $\theta$-dependent functions $\zeta_\pm$ satisfy the following differential equations

\begin{equation}
 \mathcal{D}_0 \xi _ +   = rR_ +,~~~~~~
 \Delta \mathcal{D}_0^\dag  \xi _ -   = rR_ - \label{diffeqtnxi},
\end{equation}
\begin{equation}
 \mathcal{L}_0^\dag  \zeta _ +   = S_ +  \cos \theta,  ~~~~~~ 
 \mathcal{L}_0 \zeta _ -   = \cos \theta S_ - .\label{diffeqtnvarsigma}
\end{equation}
The differential equation (\ref{PPhi0}) combined with (\ref{Phi0A}) yields the following equation
\begin{align}
	\Delta \mathcal{L}^\dag  _0 F_+  - \Delta \mathcal{D}_0 G_+  = \sqrt 2 \left( {ia\Delta R_+ \mathcal{L}^\dag  _0 \zeta_+  + S_+ \Delta \mathcal{D}_0 \xi_+ } \right)\label{FGfgplus},
\end{align}
where we have used the definitions in (\ref{FtoA}), (\ref{GtoA}), (\ref{xi}) and (\ref{zeta}).
In a similar way, the differential equation (\ref{PPhi2}) along with equation (\ref{Phi2A}) yields the following relation
\begin{align}
	\Delta \mathcal{D}^\dag  _0 G_-  + \mathcal{L}_0 \Delta F_-  =  - \sqrt 2 \left( {\Delta \mathcal{D}^\dag  _0 S_- \xi_-  + ia\mathcal{L}_0 R_- \zeta_- } \right) \label{FGfgmin}.
\end{align}
We can solve (\ref{FGfgplus}) and (\ref{FGfgmin}) to find $F_{ \pm} $ and $G_{ \pm}$ in terms of $R_{ \pm} $, $S_{ \pm} $, $\zeta_{ \pm} $ and $\xi_{ \pm} $. The solutions are given by
\begin{align}
	F_ +   = 
{{\sqrt 2 }}\left( {ia R_ +  \zeta _ +   +  \mathcal{D}_0 H_ +  } \right) \label{FtoRfp},
\end{align}
\begin{align}
	F_ -   = 
{{\sqrt 2 }}\left( { - ia R_ -  \zeta _ -   + \mathcal{D}_0^\dag  H_ -  } \right) \label{FtoRfm},
\end{align}
\begin{align}
	G_ +   = 
{{\sqrt 2 }}\left( { - \xi _ +  S_ +   + \mathcal{L}_0^\dag  H_ +  } \right) \label{GtoSgp},
\end{align}
and
\begin{align}
	G_ -   = 
{{\sqrt 2 }}\left( { - \xi _ -  S_ -   + \mathcal{L}_0 H_ -  } \right) \label{GtoSgm},
\end{align}
where $H_\pm$ are any two arbitrary functions that depend on both $r$ and $\theta$ coordinates. The presence of arbitray functions $H_\pm$ in the solutions (\ref{FtoRfp})-(\ref{GtoSgm}) is the result of identity $[\mathcal{D}_0,\mathcal{L}_0^\dag]=0$.  These functions show the freedom of Maxwell fields $A_\mu$ in the Kerr background.
Plugging (\ref{FtoRfp}) - (\ref{GtoSgm}) back to (\ref{FtoA}) and (\ref{GtoA}) provides us the general set of explicit solutions for $A_\mu$ which includes the arbitrary functions $H_\pm$. As the results are lengthy, we present them in apendix \ref{Apfullsol}. As we notice to find the solutions for $A_\mu$, we have used only the equations (\ref{Phi0F}) and (\ref{Phi2F}) for the complex scalars $\Phi_0$ and $\Phi_2$. The gauge condition is the remaining equation (\ref{Phi1F}) for $\Phi_1$. 
Using equations (\ref{eq1}) - (\ref{eq4}), one can find the following equation
\[\left( {l^\mu  n^\nu   + \bar m^\mu  m^\nu  } \right)\left( {\partial _\mu  A_\nu   - \partial _\nu  A_\mu  } \right)\]
\be
 =  - \frac{{\sqrt 2 }}
{{\left( {\bar \rho *} \right)^2 }}\left[ {\left( {\zeta_{ + } \mathcal{L}_1 S_{ +}  - \zeta_{ -} \mathcal{L}_1^\dag  S_{ -} } \right) - ia\left( {\xi_{ -} \mathcal{D}_0 R_{ -}  - \xi_{ +} \mathcal{D}_0^\dag  (\Delta R_{ +}) } \right)} \right]\label{chandra-gauge}.
\ee
Plugging the known results for $A_\mu$ in equation (\ref{Phi1F}) (that we call it as the Chandrasekhar gauge) and comparing the result with equation  (\ref{chandra-gauge}) yields the following equation that the arbitrary functions $H_\pm$ must satisfy
\be
\mathcal{D}_0^\dag  \frac{{\Delta \mathcal{D}_0 H_ +  }}
{{\left( {\bar \rho *} \right)^2 }} + \mathcal{L}_1 \frac{{\mathcal{L}_0^\dag  H_ +  }}
{{\left( {\bar \rho *} \right)^2 }} - \mathcal{D}_0 \frac{{\Delta \mathcal{D}_0^\dag  H_ -  }}
{{\left( {\bar \rho *} \right)^2 }} - \mathcal{L}_1^\dag  \frac{{\mathcal{L}_0 H_ -  }}
{{\left( {\bar \rho *} \right)^2 }} = 0\label{chandra-gaugeHH}.
\ee
The equation (\ref{chandra-gaugeHH}) imposes a constraint on the choices for the arbitrary functions $H_\pm$. 
We very roughly can compare the Chandrasekhar gauge with the
well known Lorentz gauge for the Maxwell fields in Minkowski spacetime. The 
Maxwell's equations in Minkowski spacetime are invariant under the gauge transformation $A_\mu(x)\to A_\mu(x) + \partial_\mu \Lambda(x)$ where $\Lambda(x)$ is an arbitrary function. The Lorentz gauge $\partial_\mu A^\mu = 0$ restricts the arbitrary function $\Lambda(x)$ to a function that satisfy the wave equation $\Box \Lambda(x) = 0$. The Chandrasekhar gauge (\ref{Phi1F}) resembles the Lorentz gauge. The constraint equation (\ref{chandra-gaugeHH}) for $H_\pm$ resembles to
the wave equation $\Box \Lambda(x) = 0$ for $\Lambda(x)$, where the arbitrary functions $H_\pm$ play the role of $\Lambda(x)$ 
\par
Choosing both $H_\pm$ to be zero gives the simplest solutions to the second order differential
equation (\ref{chandra-gaugeHH}). Using this choice and comparing equations (\ref{FtoRfp}), (\ref{FtoRfm}), (\ref{GtoSgp}) and (\ref{GtoSgm}) with equations (\ref{FtoA}) and (\ref{GtoA}), we get the Maxwell fields as
\begin{align}
	A_t  = \frac{{ia}}{{\rho ^2 \sqrt 2 }}\left( {\Delta R_+ \zeta_+  - R_- \zeta_-  - \sin \theta \left( {\xi_+ S_+  - \xi_- S_- } \right)} \right)\label{A_t1},
\end{align}
\begin{align}
A_r  = \frac{{ia}}{{\sqrt 2 }}\left( {R_+ \zeta_+  + \frac{{R_- \zeta_- }}{\Delta }} \right)\label{A_r1},
\end{align}
\begin{align}
A_\theta   =  - \frac{1}{{\sqrt 2 }}\left( {\xi_+ S_+  + \xi_- S_- } \right)\label{A_theta},
\end{align} and
\begin{align}
A_\phi   = \frac{{ - i}}{{\rho ^2 \sqrt 2 }}\left( {a^2 \sin ^2 \theta \left( {\Delta R_+ \zeta_+  - R_- \zeta_- } \right) - \sin \theta \left( {r^2  + a^2 } \right)\left( {\xi_+ S_+  - \xi_- S_- } \right)} \right)\label{A_phi},
\end{align}
where the functions $\zeta_\pm$ and $\xi_\pm$, given by (\ref{xi}) and (\ref{zeta}) can be rewritten as 
\be
 \xi _ +   = \frac{1}{{2CK}}\left( {\left( {ir\lambda  + 2\alpha ^2 \omega } \right)R_ -   - irC\Delta R_ +  } \right), \ee
 \be
 \xi _ -   = \frac{1}{{2CK}}\left( { - \left( {ir\lambda - 2\alpha ^2 \omega } \right)\Delta R_ +   + irCR_ -  } \right), \ee
 \be
 \zeta _ +   = \frac{1}{{2CQ}}\left( {\left( { - \lambda \cos \theta  - \frac{{2\alpha ^2 \omega }}{a}} \right)S_ -   - CS_ +  \cos \theta } \right), \ee
 \be
 \zeta _ -   = \frac{1}{{2CQ}}\left( {\left( {\lambda\cos \theta  - \frac{{2\alpha ^2 \omega }}{a}} \right)S_ +   + CS_ -  \cos \theta } \right).
\ee
The functions $K$, $Q$, and $\alpha$ are given in (\ref{Kdef}), (\ref{Qdef}), and (\ref{alpha}) respectively.  As a consistency check, we substitute the equations (\ref{A_t1})-(\ref{A_phi}) for the different components of Maxwell fields into equation $(\ref{Phi0A})$ and find $\Phi_0=R_+(r)S_+(\theta)$ in perfect agreement with what was considered in equation (\ref{PhitoTeu}) for $\Phi_0$. A similar calculation shows substituting the equations (\ref{A_t1})-(\ref{A_phi}) into equation $(\ref{Phi2A})$ yields $\Phi_2=R_-(r)S_-(\theta)$ that is again in agreement with equation (\ref{PhitoTeu}) for $\Phi_2$

\section{Boundary action for Maxwell fields in the background of Kerr black hole}
\label{sec-bound}

The action for the Maxwell fields in gravitational background $g_{\mu\nu}$ with no current is 
\begin{align}
	S = \frac{1}{4}\int {d^4 x\sqrt { - g} {\bf F^*}_{\mu \nu } {\bf F}^{\mu \nu } } + c.c. \label{action1}
\end{align}
that leads to the Maxwell's equations (\ref{eom1}) and (\ref{eom2}). The $c.c.$ term should be added in (\ref{action1}) to ensure that the action is real valued as we notice that the Chandrasekhar solutions for the Maxwell fields in Kerr spacetime  (\ref{A_t1}) - (\ref{A_phi}) are basically complex quantities. The existence of $\partial_t$ and $\partial_\phi$ Killing vectors in Kerr geometry leads to write down the dependence of Maxwell fields ${\bf A}$ on coordinates $t$ and $\phi$ as
\be
{\bf A} = 
\left( {\begin{array}{*{20}c}
   {{\bf A}_t }  \\
   {{\bf A}_r }  \\
   {{\bf A}_\theta  }  \\
   {{\bf A}_\phi  }  \\
\end{array}} \right) = e^{ - i\omega t + im\phi } \left( {\begin{array}{*{20}c}
   {A_t }  \\
   {A_r }  \\
   {A_\theta  }  \\
   {A_\phi  }  \\
\end{array}} \right),
\ee
where 
$A_\mu$'s are given by (\ref{A_t1}) - (\ref{A_phi}). We note that in (\ref{action1}), ${\bf F}_{\mu\nu} = \partial_\mu {\bf A}_\nu - \partial_\nu {\bf A}_\mu$ and so we can write
$S = 2S_0$ where 
\begin{align}
	S_0 = \frac{1}{4}\int {d^4 x\sqrt { - g} \left( {\partial _\mu  {\bf A}_\nu^*  } \right){\bf F}^{\mu \nu } }  + c.c. \label{S1}.
\end{align}
The integrand of $S_0$ can be written as the difference of a total derivative term and other term which is, in fact, proportional to the Maxwell's equations (\ref{eom2}). Taking a spherical boundary with radius $r_B$ that is the boundary of near-NHEK geometry of Kerr black hole, we can convert the total derivative term to a boundary term, given by
\begin{align}
	S_B  = \frac{1}{2}\int {d^3 x\left. {\sqrt { - g} {\bf A}_\nu^*  {\bf F}^{r\nu } } \right|_{r = r_B } }  + c.c.\label{boundary-action},
\end{align}
where 
$d^{3}x$ stands for $dt d\phi d\theta$.
The field strength tensor components
\begin{equation}
{\bf F}^{r\nu }  = g^{rr} g^{\nu \beta } {\bf F}_{r\beta }  = g^{rr} g^{\nu \beta } \left( {\partial _r  \bf A_\beta  - \partial _\beta  \bf A_r } \right),
\end{equation}
can be written simply as ${\bf{ F}}  = {\bf{ \Xi A}} $
where
\begin{align}
	{\bf{ \Xi}} = g^{rr} \left( {\begin{array}{*{20}c}
   {g^{tt} \partial _r } & { - \left( {g^{tt} \partial _t  + g^{t\phi } \partial _\phi  } \right)} & 0 & {g^{t\phi } \partial _r }  \\
   0 & 0 & 0 & 0  \\
   0 & { - g^{\theta \theta } \partial _\theta  } & {g^{\theta \theta } \partial _r } & 0  \\
   {g^{\phi t} \partial _r } & { - \left( {g^{\phi t} \partial _t  + g^{\phi \phi } \partial _\phi  } \right)} & 0 & {g^{\phi \phi } \partial _r }  \\
\end{array}} \right)\label{DD}.
\end{align}
Using the above expressions, we can rewrite the boundary action (\ref{boundary-action}) accordingly as
\begin{align}
	S_B  = \frac{1}{2}\int {d^3 x\left. {\sqrt { - g} {\bf{ A^\dag \Xi  A}}  } \right|} _{r = r_B } + c.c.,\label{Smat}
\end{align}
where
\[
{\bf{ A^\dag \Xi  A}}  = g^{rr} \left( {g^{tt} \left( {A_t^* \partial _r A_t  - i\omega A_t ^*A_r  } \right) + g^{t\phi } \left( {A_t ^*\partial _r A_\phi   + imA_t ^*A_r  } \right) - g^{\theta \theta } \left( {A_\theta ^* \partial _\theta  A_r   - A_\theta  ^*\partial _r A_\theta   } \right)} \right.
\]
\begin{align}
	\left. { + g^{\phi t} \left( {A_\phi ^* \partial _r A_t   - i\omega A_\phi ^* A_r  } \right) + g^{\phi \phi } \left( {A_\phi ^* \partial _r A_\phi   + imA_\phi  ^*A_r  } \right)} \right).\label{ADA}
\end{align}

\section{Approximations for Maxwell fields in the near horizon limit}
\label{sec-app}

The solutions for Maxwell fields in Kerr background that are given in (\ref{A_t1}) - (\ref{A_phi}) contain the radial Teukolsky functions $R_\pm(r)$.
In \cite{Hartman:2009nz}, the authors find the exact solutions to the radial Teukolsky equations for spin weight $\pm 1$ in the corresponding ``near'' region $x\ll 1$ and ``far'' region $x \gg \tau_H$  where
\begin{equation}
x = \frac{r-r_+}{r_+}\label{x},
\end{equation}
where $\tau _H  = \frac{{r_ +   - r_ -  }}{{r_ +  }}$ is the dimensionless Hawking temperature \cite{Hartman:2009nz,Becker:2012vda} which is related to the Hawking temperature $T_H$ 
of the Kerr black holes by  
\be
T_H  = \frac{{\tau _H }}{{8\pi M}}\label{THTauH}.
\ee

As we are discussing the near extremal rotating black holes, thus the dimensionless Hawking temperature $\tau_H$ would be very small number. This fact would play an important role later in getting the dominant terms of the action that describe the Maxwell fields in near horizon of near extremal Kerr black holes.
To get the solutions everywhere, the incomplete solutions from ``near'' and ``far'' regions should match in the ``matching'' region. \par
The solutions to Teukolsky radial equations (\ref{Teu1}) and (\ref{Teu3}) 
in the matching region can be read as \cite{Hartman:2009nz}
\begin{align}
	R_ +   = N_+ \tau _H^{ - 1 - in/2} \left( {\mathcal{A}_ +  \left( {\frac{r}{{\tau _H }}} \right)^{\beta  - 3/2}  + \mathcal{B}_ +  \left( {\frac{r}{{\tau _H }}} \right)^{ - \beta  - 3/2} } \right) + ...,\label{Rpsol}
\end{align}
\begin{align}
	R_ -   = N_- \tau _H^{1 - in/2} \left( {\mathcal{A}_ -  \left( {\frac{r}{{\tau _H }}} \right)^{\beta  + 1/2}  + \mathcal{B}_ -  \left( {\frac{r}{{\tau _H }}} \right)^{ - \beta  + 1/2} } \right) + ...,\label{Rmsol}
\end{align}
where $\beta$ is given by
\begin{equation}
\beta ^2  = \frac{1}{4} + K_l  - 2m^2\label{beta}.
\end{equation}
The parameter $K_l$ is related to $\lambda$ in equations (\ref{Teu1})-(\ref{Teu4}) by $K_l=\lambda+2am\omega$ and  we consider $K_l \geq 2m^2 -1/4$ and so $\beta$ is a real number. The coefficients $\mathcal{A}_\pm$ and $\mathcal{B}_\pm$ are
\begin{equation}
\mathcal{A}_\pm  = \frac{{\Gamma \left( {2\beta } \right)\Gamma \left( {1 \mp 1 - i{n}} \right)}}{{\Gamma \left( {\frac{1}{2} + \beta  - i\left( {{n} - m} \right)} \right)\Gamma \left( {\frac{1}{2} + \beta  \mp 1 - im} \right)}},\label{Acoeff}
\end{equation}
\begin{equation}
\mathcal{B}_\pm  = \frac{{\Gamma \left( { - 2\beta } \right)\Gamma \left( {1 \mp 1 - i{n}} \right)}}{{\Gamma \left( {\frac{1}{2} - \beta  - i\left( {{n} - m} \right)} \right)\Gamma \left( {\frac{1}{2} - \beta  \mp 1 - im} \right)}},\label{Bcoeff}
\end{equation}
where 
\be
{n} = \frac{\omega - m\Omega_H}{2\pi T_H},\label{neq}
\ee
and $\Omega_H=\frac{a}{r_+^2+a^2}$ is the angular velocity of the horizon. We notice that since $\tau_H$ is a very small number and $n$ is a finite number, so the equation (\ref{neq})  implies $\omega \sim m\Omega_H$. This means we consider only the Maxwell fields with frequency that is around the superradiant bound. The coefficients $N_+$ and $N_-$ are the normalization constants that their ratio can be fixed (by using equation  (\ref{RpRm})) to
\begin{equation}
\frac{{N_ -  }}{{N_ +  }} =  - \frac{{{\cal K}_lr_ + ^2 }}{{n\left( {n + i} \right)}}\label{relativeNorm}.
\end{equation}
where ${\cal K}_l=\sqrt{K_l^2+m^2(m^2+1-2K_l)}$. In deriving the ratio (\ref{relativeNorm}), we considered the near horizon limit $r \to r_+$.
As we notice from (\ref{ADA}), we need to find the derivative of the gauge fields with respect to the coordinate $r$. From equations (\ref{Rpsol}) and (\ref{Rmsol}), we find the following equations 
\begin{align}
	\partial _r R_ +   = \left( {\frac{{\beta  - 3/2}}{r}} \right)R_ +   - Q_ +,~~~ \partial _r R_ -   = \left( {\frac{{\beta  + 1/2}}{r}} \right)R_ -   - Q_ -, \label{IDR}
\end{align}
where \begin{equation}
Q_ +   \equiv \frac{{2\beta \mathcal{B}_ +}}{r}\tau _H^{ - 1 - in/2} \left( {\frac{r}{{\tau _H }}} \right)^{ - \beta  - 3/2},~~ Q_ -   \equiv \frac{{2\beta \mathcal{B}_ -}}{r}\tau _H^{1 - in/2} \left( {\frac{r}{{\tau _H }}} \right)^{ - \beta  + 1/2} \label{Qpm}.
\end{equation}
As we notice from expressions (\ref{A_t1}) and (\ref{A_phi}),  the gauge field components $A_t$ and $A_\phi$ depend on coordinates $r$ and $\theta$ in a non-separable way, due to the presence of function $\rho=\sqrt{r^2+a^2\cos^2\theta}$. As a result, performing the integration over the boundary in (\ref{boundary-action}) becomes almost impossible. 
We can separate the dependence of gauge fields (\ref{A_t1}) and (\ref{A_phi}) on $r$ and $\theta$ by making an approximation. The approximation is to set the coordinate $r$ equal to the boundary radius $r_B$ in all  $\rho$ and $\Delta$ that appear in (\ref{A_t1}) - (\ref{A_phi}). In this approximation, $\Delta_B=\Delta(r=r_B)=(r_B-r_+)(r_B-r_-)$ approaches to zero as the boundary radius $r_B\rightarrow r_+$.  So, we can write the Maxwell fields on the boundary as
\begin{eqnarray}
 {\bf A}_t  &=& e^{ - i\omega t + im\phi } \left( {\left( {f_1 S_ -   + f_2 S_ +  } \right)R_ +  \Delta_B  + \left( {f_3 S_ -   + f_4 S_ +  } \right)R_ -  } \right),\label{Adef1} \\ 
 {\bf A}_r  &=& e^{ - i\omega t + im\phi } \left( {\left( {f_5 S_ -   + f_6 S_ +  } \right)R_ +   + \left( {f_7 S_ -   + f_8 S_ +  } \right)R_ -  \Delta_B ^{ - 1} } \right),\label{Adef2} \\ 
 {\bf A}_\theta  &=& e^{ - i\omega t + im\phi } \left( {\left( {f_9 S_ -   + f_{10} S_ +  } \right)R_ +  \Delta_B  + \left( {f_{11} S_ -   + f_{12} S_ +  } \right)R_ -  } \right),\label{Adef3} \\ 
 {\bf A}_\phi   &=& e^{ - i\omega t + im\phi } \left( {\left( {f_{13} S_ -   + f_{14} S_ +  } \right)R_ +  \Delta_B  + \left( {f_{15} S_ -   + f_{16} S_ +  } \right)R_ -  } \right), \label{Adef4}
 \end{eqnarray}
where $f_i$'s ($i = 1,...,16$) depend only on $\theta$ and are given by
\begin{equation}
f_1 = \frac{{ia\sqrt 2  {\left( { - \lambda \cos \theta  - 2\alpha ^2 \omega a^{ - 1} } \right) } }}{{4\rho _B^2 C\left( { - a\omega \sin \theta  + m\left( {\sin \theta } \right)^{ - 1} } \right)}} - \frac{{ia\sqrt 2 \sin \theta  { \left( {ir_B \lambda  - 2\alpha ^2 \omega } \right)} }}{{4C\left( {am - \omega \left( {r_B^2  + a^2 } \right)} \right)\rho _B^2 }},\label{f1}
\end{equation}\begin{equation}
f_2 = -\frac{{ia\sqrt 2   \cos \theta}}{{4\rho _B^2 \left( { - a\omega \sin \theta  + m\left( {\sin \theta } \right)^{ - 1} } \right)}} - \frac{{a r_B\sqrt 2 \sin \theta}}{{4\left( {am - \omega \left( {r_B^2  + a^2 } \right)} \right)\rho _B^2 }},
\end{equation}\begin{equation}
f_3 = - \frac{{ia\sqrt 2  \cos \theta}}{{4\rho _B^2 \left( { - a\omega \sin \theta  + m\left( {\sin \theta } \right)^{ - 1} } \right)}} - \frac{{ar_B\sqrt 2 \sin \theta }}{{4\left( {am - \omega \left( {r_B^2  + a^2 } \right)} \right)\rho _B^2 }},
\end{equation}\begin{equation}
f_4 = - \frac{{ia\sqrt 2 \left( {\lambda \cos \theta  - 2\alpha ^2 \omega a^{ - 1} } \right)}}{{4\rho _B^2 C\left( { - a\omega \sin \theta  + m\left( {\sin \theta } \right)^{ - 1} } \right)}} - \frac{{ia\sqrt 2 \sin \theta \left( {ir_B \lambda  + 2\alpha ^2 \omega } \right)}}{{4C\left( {am - \omega \left( {r_B^2  + a^2 } \right)} \right)\rho _B^2 }},
\end{equation}\begin{equation}
f_5 = \frac{{ia\sqrt 2 \left( { - \lambda \cos \theta  - 2\alpha ^2 \omega a^{ - 1} } \right)}}{{4C\left( { - a\omega \sin \theta  + m\left( {\sin \theta } \right)^{ - 1} } \right)}},
\end{equation}\begin{equation}
f_6  = -\frac{{ia\sqrt 2 \cos \theta}}{{4\left( { - a\omega \sin \theta  + m\left( {\sin \theta } \right)^{ - 1} } \right)}},
\end{equation}\begin{equation}
f_7  = \frac{{ia\sqrt 2 \cos \theta}}{{4\left( { - a\omega \sin \theta  + m\left( {\sin \theta } \right)^{ - 1} } \right)}},
\end{equation}\begin{equation}
f_8  = \frac{{ia\sqrt 2 \left( {\lambda \cos \theta  - 2\alpha ^2 \omega a^{ - 1} } \right)}}{{4C\left( { - a\omega \sin \theta  + m\left( {\sin \theta } \right)^{ - 1} } \right)}},
\end{equation}\begin{equation}
f_{9}  =  \frac{{\sqrt 2  \left( {ir_B \lambda  - 2\alpha ^2 \omega } \right)}}{{4C\left( {am - \omega \left( {r_B^2  + a^2 } \right)} \right)}},
\end{equation}\begin{equation}
f_{10}  =  \frac{{i r_B\sqrt 2 }}{{4\left( {am - \omega \left( {r_B^2  + a^2 } \right)} \right)}},
\end{equation}\begin{equation}
f_{11}  = -\frac{{ir_B \sqrt 2 }}{{4\left( {am - \omega \left( {r_B^2  + a^2 } \right)} \right)}},
\end{equation}\begin{equation}
f_{12}  = -\frac{{\sqrt 2 \left( {ir_B \lambda - 2\alpha ^2 \omega } \right)}}{{4C\left( {am - \omega \left( {r_B^2  + a^2 } \right)} \right)}},
\end{equation}\begin{equation}
f_{13}  = \frac{{ia^2 \sqrt 2 \left( {\lambda \cos \theta  + 2\alpha ^2 \omega a^{ - 1} } \right)\sin ^2 \theta }}{{4\rho _B^2 C\left( { - a\omega \sin \theta  + m\left( {\sin \theta } \right)^{ - 1} } \right)}}
+ \frac{{i\left( {r_B^2  + a^2 } \right)\sqrt 2 \sin \theta \left( {ir_B \lambda  - 2\alpha ^2 \omega } \right)}}{{4C\left( {am - \omega \left( {r_B^2  + a^2 } \right)} \right)\rho _B^2 }},
\end{equation}\begin{equation}
f_{14}  = \frac{{ia^2 \sqrt 2 \cos \theta \sin ^2 \theta }}{{4\rho _B^2 \left( { - a\omega \sin \theta  + m\left( {\sin \theta } \right)^{ - 1} } \right)}}
+ \frac{{r_B\left( {r_B^2  + a^2 } \right)\sqrt 2 \sin \theta }}{{4\left( {am - \omega \left( {r_B^2  + a^2 } \right)} \right)\rho _B^2 }},
\end{equation}\begin{equation}
f_{15}  = \frac{{ia^2 \sqrt 2 \cos\theta\sin ^2 \theta }}{{4\rho _B^2 \left( { - a\omega \sin \theta  + m\left( {\sin \theta } \right)^{ - 1} } \right)}} + \frac{{r_B \left( {r_B^2  + a^2 } \right)\sqrt 2 \sin \theta }}{{4\left( {am - \omega \left( {r_B^2  + a^2 } \right)} \right)\rho _B^2 }},
\end{equation}\begin{equation}
f_{16}  = \frac{{ia^2 \sqrt 2 \left( {\lambda\cos \theta  - 2\alpha ^2 \omega a^{ - 1} } \right)\sin ^2 \theta }}{{4\rho _B^2 C\left( { - a\omega \sin \theta  + m\left( {\sin \theta } \right)^{ - 1} } \right)}} + \frac{{i\left( {r_B^2  + a^2 } \right)\sqrt 2 \sin \theta \left( {ir_B \lambda  - 2\alpha ^2 \omega } \right) }}{{4C\left( {am - \omega \left( {r_B^2  + a^2 } \right)} \right)\rho _B^2 }}.
\label{f16}
\end{equation}
In (\ref{f1}) - (\ref{f16}), 
\be
\rho_B = r_B^2 + a^2 \cos^2 \theta \label{rhoB}.
\ee
We note that for the near extremal Kerr black holes $\Omega_H \simeq \frac{1}{2a}$ and so $\omega \sim \frac{m}{2a}$.  As a result for $r_B \rightarrow r_+$, all $f_i$'s (except $f_5,\cdots,f_8$ that appear in radial component (\ref{Adef2}) of Maxwell field) become very large. Moreover due to
smallness of $\Delta_B$ in the near horizon limit of near extremal black hole, all components 
of Maxwell fields in the near horizon of near extremal Kerr background become very large. We consider the difference between $am$ and $\omega(r_B^2+a^2)$ in near horizon near extremal Kerr black hole to be the same order of  $\Delta_B$.
\par

\section{Two-point function of vector fields}
\label{sec-2ptvarMax}

In this section we explicitly calculate the boundary action (\ref{boundary-action}) and find the two-point function of the vector fields. We rewrite the 
components of the Maxwell field {(\ref{Adef1}) - (\ref{Adef4})} in a matrix form as
\be
{\bf{ A}}  = e^{ - i\omega t + im\phi } (R_ +  {\bf{ K  v}}_ +   + R_ - {\bf{ L  v}}_ -) , \label{Amatrixpm}
\ee
where the matrices $\bf{K}$ and $\bf{L}$ are
\be
{\bf{ K}} = \left( {\begin{array}{*{20}c}
   {\Delta_B f_1 } & 0 & 0 & {\Delta_B f_2 }  \\
   {f_5 } & {\kappa _1 } & 0 & {f_6 }  \\
   {\Delta_B f_9 } & 0 & {\kappa _2 } & {\Delta_B f_{10} }  \\
   {\Delta_B f_{13} } & 0 & 0 & {\Delta_B f_{14} }  \\
\end{array}} \right) ,~~~ {\bf{ L}} = \left( {\begin{array}{*{20}c}
   {\kappa _3 } & {f_3 } & {f_4 } & 0  \\
   0 & {f_7 \Delta_B ^{ - 1} } & {f_8 \Delta_B ^{ - 1} } & 0  \\
   0 & {f_{11} } & {f_{12} } & 0  \\
   0 & {f_{15} } & {f_{16} } & {\kappa _4 }  \\
\end{array}} \right)\label{KL},
\ee
and
\be
{\bf{ v}}_ +   = \left( {\begin{array}{*{20}c}
   {S_ -  } & 0 & 0 & {S_ +  }  \\
\end{array}} \right)^T,~~~ {\bf{ v}}_ -   = \left( {\begin{array}{*{20}c}
   0 & {S_ -  } & {S_ +  } & 0  \\
\end{array}} \right)^T \label{vpvm}.
\ee
For later convenience, we show the first and the second term of (\ref{Amatrixpm}) by 
${\bf{ A}}_ +$ and ${\bf{ A}}_ -$, respectively. 
We notice the matrices ${\bf K}$ and ${\bf L}$ as well as vectors ${\bf{v}}_\pm$ depend only on angular coordinate $\theta$, according to Teukolsky equations (\ref{Teu2}), (\ref{Teu4}) and equations (\ref{f1}) - (\ref{f16}) for $f_i$. 
The arbitrary constants $\kappa_i$, $i = 1,2,3,4$ in (\ref{KL}) are introduced to provide the invertibility for matrices ${\bf K}$ and ${\bf L}$.  However we notice that $\kappa _i \rightarrow 0$ to reduce the Maxwell fields (\ref{Amatrixpm}) to the solutions {(\ref{Adef1}) - (\ref{Adef4})} and we perform this limit at the end of calculation wherever $\kappa_i$'s appear.
We may find that due to the gauge choice (\ref{Phi1F}), there is a relation between the vectors $\bf{ v}_+$ and $\bf{ v}_-$ as $\bf{ v}_- = \chi\bf{ v}_+$ where the matrix $\chi$ is given by
\begin{equation}
\chi  = \left( {\begin{array}{*{20}c}
   0 & 0 & 0 & 0  \\
   1 & 0 & 0 & 0  \\
   0 & 0 & 0& 1  \\
   0 & 0 & 0 & 0\\
\end{array}} \right).\label{chimatrix1}
\end{equation}
Denoting the Maxwell fields and the Teukolsky functions on the boundary by ${\bf A}_\pm^B $ and $R_\pm^B$ respectively,  
we get 
\begin{equation}
\frac{ {{\bf{ { A}}}_{_ +  }^B } }{{R_ + ^B }} = e^{ - i\omega t + im\phi } {\bf{ K  v}}_ +  \label{Abulk2},
\end{equation}
and
\begin{equation}
\frac{ {{\bf{{ A}}}_{_ -  }^B } }{{R_ - ^B }} = e^{ - i\omega t + im\phi } {\bf{ L v}}_ -  \label{Abulk4}.
\end{equation}
As we notice, equations (\ref{Abulk2}) and (\ref{Abulk4}) enable us to consider the non-radial dependent parts of Maxwell fields as the ratio of boundary Maxwell fields to the boundary Teukolsky radial solutions. 
Using the relation between $\bf{ v}_+$ and $\bf{ v}_-$, we have
\begin{equation}
\begin{array}{l}
 {\bf{ A}}  
  = \left( {R_ +  {\bf{K}} + R_ -  {\bf{L\chi }}} \right)e^{ - i\omega t + im\phi } {\bf{v}}_ +   = \left( {R_ +   + R_ -  {\bf{L\chi K}}^{{\bf{ - 1}}} } \right)\frac{{\left( {{\bf{ A}}_ + ^B } \right)  }}{{R_ + ^B }} \\ 
 \end{array}\label{Abulk5},
\end{equation}
or
\begin{equation}
{\bf{A}}^\dag = 
\frac{{\left( {{\bf{A}}_ + ^B } \right)^\dag}}{{\left( {R_ + ^B } \right)^* }}\left( {R_ + ^*  + R_ - ^* \left( {{\bf{L\chi K}}^{{\bf{ - 1}}} } \right)^\dag  } \right) \label{Abulk6}.
\end{equation}
\par
We calculate the integrand  $\bf {A}^{\dag}\Xi \bf{A}$ of the boundary action (\ref{boundary-action}) now. We notice that though the matrix $\bf{\Xi}$ in (\ref{DD}) has real entries, but after acting on $\bf{A}$, the result is a complex-valued vector. 
We decompose the operator  $\bf{\Xi}$ as
\begin{equation}
{\bf{\Xi}} = g^{rr}\left(\bf{\Pi} + \bf{\Theta} \right),
\end{equation}
where
\begin{equation}
{\bf{\Pi}}  = \left( {\begin{array}{*{20}c}
   {g^{tt} \partial _r } & 0 & 0 & {g^{t\phi } \partial _r }  \\
   0 & 0 & 0 & 0  \\
   0 & 0 & {g^{\theta \theta } \partial _r } & 0  \\
   {g^{\phi t} \partial _r } & 0 & 0 & {g^{\phi \phi } \partial _r }  \\
\end{array}} \right),
\end{equation}
contains only the derivatives with respect to $r$ and
\begin{equation}
{\bf{\Theta}}  = \left( {\begin{array}{*{20}c}
   0 & { - \left( {g^{tt} \partial _t  + g^{t\phi } \partial _\phi  } \right)} & 0 & 0  \\
   0 & 0 & 0 & 0  \\
   0 & { - g^{\theta \theta } \partial _\theta  } & 0 & 0  \\
   0 & { - \left( {g^{\phi t} \partial _t  + g^{\phi \phi } \partial _\phi  } \right)} & 0 & 0  \\
\end{array}} \right),
\end{equation}
contains the derivatives with respect to $t$ and $\phi$.
The reason for performing this decomposition is due to the fact that the radial dependence of the Maxwell fields (\ref{Adef1}) - (\ref{Adef4})  
are in terms of functions $R_\pm$, while the non-radial dependence are in 
$\frac{{{\bf{A}}_ + ^B }}{R_+^B}$ or $\frac{{{\bf{A}}_ - ^B }}{R_-^B}$ (according to (\ref{Abulk2}) and (\ref{Abulk4})). 
We find $\bf {A}^{\dag}\Xi \bf{A}$ is given by
\begin{eqnarray}
 {\bf{A^\dag\Xi A}}   &=& g^{rr} \frac{{\left( {{\bf{A}}_ + ^B } \right)^\dag  }}{{\left( {R_ + ^B } \right)^* }}\left( {R_ +   + R_ -  {\bf{L\chi K}}^{ - 1} } \right)^\dag  \left({\bf{\Pi +\Theta }}\right)\left( {R_ +   + R_ -  {\bf{L\chi K}}^{ - 1} } \right)\frac{{{\bf{A}}_ + ^B }}{{R_ + ^B }} \nonumber \\ 
&=& g^{rr} \frac{{\left( {{\bf{A}}_ + ^B } \right)^\dag  }}{{\left( {R_ + ^B } \right)^* }}R_ + ^* \left( {{\bf{\Pi }}R_ + + R_+ {\bf{\Theta}} } \right)\frac{{{\bf{A}}_ + ^B }}{{R_ + ^B }} 
+ g^{rr} \frac{{\left( {{\bf{A}}_ + ^B } \right)^\dag  }}{{\left( {R_ + ^B } \right)^* }}R_ + ^* \left( {{\bf{\Pi }}R_ - + R_- {\bf{\Theta}} } \right){\bf{L\chi K}}^{ - 1} \frac{{{\bf{A}}_ + ^B }}{{R_ + ^B }} \nonumber\\
&+& g^{rr} \frac{{\left( {{\bf{A}}_ + ^B } \right)^\dag  }}{{\left( {R_ + ^B } \right)^* }}\left( {R_ -  {\bf{L\chi K}}^{ - 1} } \right)^\dag  \left( {{\bf{\Pi }}R_ + + R_+ {\bf{\Theta}} } \right)\frac{{{\bf{A}}_ + ^B }}{{R_ + ^B }}\nonumber\\
&+& g^{rr} \frac{{\left( {{\bf{A}}_ + ^B } \right)^\dag  }}{{\left( {R_ + ^B } \right)^* }}\left( {R_ -  {\bf{L\chi K}}^{ - 1} } \right)^\dag  \left( {{\bf{\Pi }}R_ - + R_- {\bf{\Theta}} } \right){\bf{L\chi K}}^{ - 1} \frac{{{\bf{A}}_ + ^B }}{{R_ + ^B }} . \label{AChiA0} 
 \end{eqnarray}
As the matrix $\bf{\Pi}$ contains the differential operator $\partial _r$, it would be helpful to split ${{\bf{\Pi }}R_ +  }$,
\[
{\bf{ \Pi}}R_ +   = \left( {\begin{array}{*{20}c}
   {g^{tt} \partial _r R_ +  } & 0 & 0 & {g^{t\phi } \partial _r R_ +  }  \\
   0 & 0 & 0 & 0  \\
   0 & 0 & {g^{\theta \theta } \partial _r R_ +  } & 0  \\
   {g^{\phi t} \partial _r R_ +  } & 0 & 0 & {g^{\phi \phi } \partial _r R_ +  }  \\
\end{array}} \right)
\]
\begin{equation}
 = \left( {\begin{array}{*{20}c}
   {g^{tt} \left( {\left( {\frac{{\beta  - 3/2}}{{r}}} \right)R_ +   - Q_ +  } \right)} & 0 & 0 & {g^{t\phi } \left( {\left( {\frac{{\beta  - 3/2}}{{r}}} \right)R_ +   - Q_ +  } \right)}  \\
   0 & 0 & 0 & 0  \\
   0 & 0 & {g^{\theta \theta } \left( {\left( {\frac{{\beta  - 3/2}}{{r}}} \right)R_ +   - Q_ +  } \right)} & 0  \\
   {g^{\phi t} \left( {\left( {\frac{{\beta  - 3/2}}{{r}}} \right)R_ +   - Q_ +  } \right)} & 0 & 0 & {g^{\phi \phi } \left( {\left( {\frac{{\beta  - 3/2}}{{r}}} \right)R_ +   - Q_ +  } \right)}, \\
\end{array}} \right),
\end{equation}
to two terms, given by
\be
  {\bf{ \Pi}}R_ + =R_ +  {\bf{ \Pi}}_1  - Q_ +  {\bf{ \Pi}}_2. \label{xiRp}
\ee
In (\ref{xiRp}), the matrix ${\bf{ \Pi}}_2$ is given by
\be
{\bf{ \Pi}}_2  = \left( {\begin{array}{*{20}c}
   {g^{tt} } & 0 & 0 & {g^{t\phi } }  \\
   0 & 0 & 0 & 0  \\
   0 & 0 & {g^{\theta \theta } } & 0  \\
   {g^{\phi t} } & 0 & 0 & {g^{\phi \phi } }  \\
\end{array}} \right)\label{xi2},
\ee
and ${\bf{ \Pi}}_1={\frac{{\beta  - 3/2}}{{r}}}{\bf{ \Pi}}_2$. A similar calculation shows we can split ${{\bf{\Pi }}R_ -  }$ to two terms, 
as
\be
 {{\bf{\Pi }}R_ -  }=R_ -  {\bf{ \Pi}}_3  - Q_ -  {\bf{ \Pi}}_4, \label{xiRm}
\ee
where ${\bf{ \Pi}}_3={\frac{{\beta  + 1/2}}{{r}}}{\bf{ \Pi}}_2$ and ${\bf{ \Pi}}_4={\bf{ \Pi}}_2$.
So, in terms of  ${\bf{\Pi}}_1$, ${\bf{\Pi}}_2$, ${\bf{\Pi}}_3$ and ${\bf{\Pi}}_4$, we get the following expression for (\ref{AChiA0})
\begin{eqnarray}
{\bf{A^\dag\Xi A}}  & =& g^{rr} \frac{{\left( {{\bf{A}}_ + ^B } \right)^\dag  }}{{\left( {R_ + ^B } \right)^* }}R_ + ^* \left( {R_ +  ({\bf{\Pi }}_1 +{\bf{\Theta}})   - Q_ +  {\bf{\Pi }}_2 } \right)\frac{{{\bf{A}}_ + ^B }}{{R_ + ^B }}
\nonumber\\
&+& g^{rr} \frac{{\left( {{\bf{A}}_ + ^B } \right)^\dag  }}{{\left( {R_ + ^B } \right)^* }}R_ + ^* \left( {R_ -  ({\bf{\Pi }}_3 +{\bf{\Theta}})  - Q_ -  {\bf{\Pi }}_4 } \right){\bf{L\chi K}}^{ - 1} \frac{{{\bf{A}}_ + ^B }}{{R_ + ^B }} \nonumber\\ 
 &+& g^{rr} \frac{{\left( {{\bf{A}}_ + ^B } \right)^\dag  }}{{\left( {R_ + ^B } \right)^* }}\left( {R_ -  {\bf{L\chi K}}^{ - 1} } \right)^\dag  \left( {R_ +  ({\bf{\Pi }}_1 +{\bf{\Theta}})  - Q_ +  {\bf{\Pi }}_2 } \right)\frac{{{\bf{A}}_ + ^B }}{{R_ + ^B }}\nonumber\\
&+& g^{rr} \frac{{\left( {{\bf{A}}_ + ^B } \right)^\dag  }}{{\left( {R_ + ^B } \right)^* }}\left( {R_ -  {\bf{L\chi K}}^{ - 1} } \right)^\dag  \left( {R_ -  ({\bf{\Pi }}_3 +{\bf{\Theta}})  - Q_ -  {\bf{\Pi }}_4 } \right){\bf{L\chi K}}^{ - 1} \frac{{{\bf{A}}_ + ^B }}{{R_ + ^B }}. \label{ADAcal}
 \end{eqnarray}
Comparing the functions $Q_\pm$ in (\ref{Qpm}) to the leading terms of $R_\pm$ in (\ref{Rpsol}) and (\ref{Rmsol}), we find $Q_\pm \sim \tau_H^{2\beta} R_\pm$. So, we can neglect the terms that are proportional to $Q_\pm$ compared to the terms that are proportional to $R_\pm$ in (\ref{ADAcal}). This yields equation (\ref{AAcalnoQ}) in appendix \ref{ap3}.
We calculate explicitly and present all the terms of (\ref{AAcalnoQ}) in the near region limit where 
$\Delta_B \ll 1$. We find that the leading terms in (\ref{ADAcal}) (or (\ref{AAcalnoQ})) are the terms that contain $\left( {{\bf{L\chi K}}^{ - 1} } \right)^\dag {\bf{\Pi _3}}  {\bf{L\chi K}}^{ - 1} $ and
$\left( {{\bf{L\chi K}}^{ - 1} } \right)^\dag {\bf{\Theta}}  {\bf{L\chi K}}^{ - 1} $ respectively. Both terms are in the order of $\Delta_B^{-3}$ as the contravariant components of the metric tensor,  $g^{tt}~,g^{t\phi}$ and $g^{\phi\phi}$, are in order of $\Delta_B^{-1}$. 
Using the results of  appendix \ref{ap3}, the boundary action (\ref{Smat}) turns out to be
\be
S_B  = \frac{1}{2}\int {dtd\theta d\phi {\sqrt { - g_B} g^{rr}(r_B) \frac{{(R_ - ^{B})^* R_ -^B  }}{{\left( {R_ + ^B } \right)^* R_ + ^B }}\tilde \theta _4^{ij}(\theta) A_{i+}^{B*}(t,\theta,\phi) A_{j+}^B  }(t,\theta,\phi)} + c.c.\label{Sbleading}.
\ee
where $\tilde \theta _4^{ij}$ is given by (\ref{tildetheta}). We show the $(t,\phi)$-dependence of the boundary gauge fields $A_{i+}^B$ by $a_{i+}^B$ according to
{
\be
A_{i + }^B = a_{i + }^B (t,\phi){\tilde{\theta}}_i(\theta)\label{sepA},
\ee
where 
\be
a_{i + } ^B = e^{-i\omega t + im\phi} R_ +^B  \Delta_B,\label{ainor}
\ee
for $i = t,\theta,\phi$ and
\be
a_{r + }^B = e^{-i\omega t + im\phi} R_ +^B. \label{ar}
\ee  
The $\theta$-dependent functions ${\tilde{\theta}}_i(\theta)$ are given by
\be
\begin{array}{l}
 {\tilde{\theta}}_t   =  {f_1 S_ -   + f_2 S_ +  },  \\ 
 {\tilde{\theta}}_r   =  {f_5 S_ -   + f_6 S_ +  },  \\ 
 {\tilde{\theta}}_\theta   =  {f_9 S_ -   + f_{10} S_ +  }, \\ 
 {\tilde{\theta}}_\phi  =  {f_{13} S_ -   + f_{14} S_ +  }. \\ 
 \end{array}\label{tildetheta}
\ee
}
Taking the functional derivative of (\ref{Sbleading}) with respect to the real part of the rescaled boundary gauge fields ${\cal A}_{i+}^B=\frac{\Re(a_{i+}^B)}{r_B^{\beta-2}\Delta_B^{3/2}}$, we get the two-point function as {\footnote {The rescaling of the boundary gauge fields is quite similar to the rescaling of the boundary gauge fields in the context of AdS/CFT correspondence \cite{mueck1}.}}
\begin{equation}
	\frac{{\delta ^2 S_B }}{{\delta{\cal A}_{i+}^B\delta {\cal A}_{j+}^B }}  
= r_B^{2\beta-4}\frac{{\left( {R_ - ^B } \right)^* R_ - ^B }}{{\left( {R_ + ^B } \right)^* R_ + ^B }} {\cal{Z}}^{ij},\label{2pt}
\end{equation}
where 
\be
{\cal{Z}}^{ij} = \int_0^\pi{d\theta \sin(\theta) 
\{\tilde\theta _4^{ij}\tilde \theta_i\tilde \theta_j^*+\tilde \theta _4^{*ij}\tilde \theta_i^*\tilde \theta_j\}}_{n.s.}\label{thetaint}.
\ee
In (\ref{thetaint}), n.s. means there is no summation over indices $i$ and $j$. Moreover we note from the results of appendix \ref{ap3} that the leading terms in (\ref{2pt}) correspond to indices $i$ and $j$ to be $t$ and $\phi$ only. 
This is an interesting result that confirms the dual CFT to four-dimensional Kerr black hole is a two-dimensional theory, in contrast to AdS/CFT correspondence that the dimension of dual CFT  always is one dimension less than the dimension of the bulk theory.
Although it looks very unlikely to perform the integration in (\ref{thetaint}) and find an exact analytical expression for ${\cal{Z}}^{ij}$, however we can find the retarded Green's function for the spin-1 fields from the factor $\frac{{\left( {R_ - ^B } \right)^* R_ - ^B }}{{\left( {R_ + ^B } \right)^* R_ + ^B }}$ in (\ref{2pt}).  
The term $\frac{{\left( {R_ - ^B } \right)^* R_ - ^B }}{{\left( {R_ + ^B } \right)^* R_ + ^B }}$ can be calculated explicitly by using the equations (\ref{Rpsol}) and (\ref{Rmsol}) as
\begin{equation}
\frac{{\left( {R_ - ^B } \right)^* R_ - ^B }}
{{\left( {R_ + ^B } \right)^* R_ + ^B }} = \left| {\frac{{N_ -  }}
{{N_ +  }}} \right|^2 \frac{{r_B^4 }}
{{{\cal A}_ + {\cal  A}_ + ^* }}\left( {{\cal  A}_ -{\cal  A}_ - ^*  + \left( {\frac{{\tau _H }}
{{r_B }}} \right)^{2\beta } \left( {{\cal  A}_ -  {\cal  B}_ - ^*  + {\cal  B}_ - {\cal   A}_ - ^* } \right) + \left( {\frac{{\tau _H }}
{{r_B }}} \right)^{4\beta } {\cal  B}_ - {\cal  B}_ - ^* } \right),\label{termGreen}
\end{equation}\nopagebreak
and so the two-point function (\ref{2pt}) becomes
\begin{eqnarray}
\frac{{\delta ^2 S_B }}{{\delta{\cal A}_{i+}^B\delta {\cal A}_{j+}^B }} &=& {\cal{Z}}^{ij}\frac{{N_ -  N_ - ^* r_B^{2\beta} }}
{{N_ +  N_ + ^* }}\left( {\left| {\frac{{{\cal  A}_ -  }}
{{{\cal  A}_ +  }}} \right|^2  + \left( {\frac{{\tau _H }}
{{r_B }}} \right)^{2\beta } \left( {\frac{{{\cal  A}_ - {\cal  B}_ - ^* }}
{{{\cal  A}_ + {\cal  A}_ + ^* }} + \frac{{{\cal  B}_ - {\cal  A}_ - ^* }}
{{{\cal  A}_ +  {\cal  A}_ + ^* }}} \right) + \left( {\frac{{\tau _H }}
{{r_B }}} \right)^{4\beta } {\cal  B}_ -  {\cal  B}_ - ^* } \right)\nonumber\\
&=&{\cal{Z}}^{ij}r_B^{2\beta}\left( M^4
+ \frac{{N_ -  N_ - ^*  }}
{{N_ +  N_ + ^* }}(\frac{\tau _H}{r_B})
 ^{2\beta } \frac{{{\cal  A}_ -  {\cal  B}_ - ^* }}
{{{\cal  A}_ +  {\cal  A}_ + ^* }} 
+ \frac{{N_ -  N_ - ^*  }}
{{N_ +  N_ + ^*  }}(\frac{\tau _H}{r_B})
^{2\beta } \frac{{{\cal  B}_ -  {\cal  A}_ - ^* }}
{{{\cal  A}_ +  {\cal  A}_ + ^* }} + \frac{{N_ -  N_ - ^* }}
{{N_ +  N_ + ^* }}{(\frac{{\tau _H }}
{{r_B }})^{4\beta}} {\cal  B}_ -  {\cal  B}_ - ^* \right).\nonumber \\ \label{diffe}
\end{eqnarray}
We have used equations (\ref{Acoeff}) and (\ref{relativeNorm}) to simplify the first term of (\ref{diffe}). Plugging for the ratios $N_-^*/N_+^*$ and ${\cal B}_-^*/{\cal A}_+^*$ that appear in the second term of (\ref{diffe}) as well as the ratios $N_-/N_+$ and ${\cal B}_-/{\cal A}_+$ in the third term from equations (\ref{Acoeff}), (\ref{Bcoeff}) and (\ref{relativeNorm}), we find
\begin{eqnarray}
\frac{{\delta ^2 S_B }}{{\delta{\cal A}_{i+}^B\delta {\cal A}_{j+}^B }}&=&{\cal{Z}}^{ij} r_B^{2\beta}\left(M^{4}
+ \left( {\frac{{{\cal K}_l M^2 }}
{{n\left( {n - i} \right)}}} \right)r_B^{-2\beta}\frac{{N_ -  {\cal A}_ -  }}
{{N_ +  {\cal A}_ +  }}G_R^*\right.
\nonumber\\
& &\,\,\,\,\,\,\,\,\,\left.
 + \left( {\frac{{{\cal K}_l M^2 }}
{{n\left( {n + i} \right)}}} \right)r_B^{-2\beta}\frac{{N_ - ^* {\cal A}_ - ^* }}
{{N_ + ^* {\cal A}_ + ^* }}G_R  + \frac{{N_ -  N_ - ^*  }}
{{N_ +  N_ + ^* }} {\frac{{\tau _H^{4\beta} }}
{{r_B^{4\beta}}}} {\cal  B}_ -  {\cal  B}_ - ^* \label{varSol}\right).
\end{eqnarray}
In (\ref{varSol}), $G_R$ stands for  
\begin{align}
	G_R(n_L,n_R)  = - n\left( {i + n} \right)T_R^{2\beta } \frac{{\Gamma \left( { - 2\beta } \right)}}
{{\Gamma \left( {2\beta } \right)}}\frac{{\Gamma \left( {\beta  + \tfrac{1}
{2} - in_R } \right)\Gamma \left( {\beta  - \tfrac{1}
{2} - in_L } \right)}}
{{\Gamma \left( {\tfrac{1}
{2} - \beta  - in_R } \right)\Gamma \left( {\tfrac{3}
{2} - \beta  - in_L } \right)}}\label{GreenF},
\end{align}
where $n_L$ and $n_R$ are related to $m$ and $\omega$ by $m = n_L$, and $n = n_L + n_R$ and we have considered the normalization (\ref{relativeNorm}) as well as the relation between dimensionless Hawking temperature $\tau_H$ with the right temperature $T_R$
\be
T_R = \frac{\tau_H}{4M\lambda}\label{Trtotau},
\ee
where $\lambda \rightarrow 0$. 
The first term in bracket in (\ref{varSol}) clearly is a constant term compared  to the other terms. The second term in (\ref{varSol}) is the complex conjugate of the third term. Moreover, we can ignore the fourth term of (\ref{varSol}) compared to the other terms, as this term is proportional to $\tau_H^{4\beta}$.
Dropping the complex conjugate term in (\ref{varSol}) according to  \cite{Becker:2012vda}, \cite{son} ,  we find that the field theoretical two-point function (\ref{2pt}) is equal to $ G_R{\cal Z}^{ij}$ up to a multiplicative  factor that depends on momentum and is not a part of the retarded Green's function. The existence of multiplicative factor has also been found for the field theoretical two-point function of spinors \cite{Becker:2012vda}. We note that  $G_R(n_L,n_R)$ (given in (\ref{GreenF})) is in exact agreement with the proposed retarded Green's function for the spin-1 fields in reference \cite{Chen:2010ni}. 
Using the optical theorem for the obtained retarded Green's function (\ref{GreenF}), we get exactly the absorption cross section of spin-1 fields scattered off of the Kerr black hole \cite{Bredberg:2009pv}.   
Interestingly enough, as we mentioned before, the boundary vector field components that contribute to the leading term of two-point function are only ${\cal A}_{t + }^B$ and ${\cal A}_{\phi + }^B$. This fact is in agreement with the statement of Kerr/CFT correspondence that the dual boundary theory is a 
two-dimensional CFT. The two-point function (\ref{2pt}) is a function of $\omega$ and $m$ which are the conjugate  momenta in $t$ and $\phi$ directions, respectively. 

\section{Two-point function of vector fields in CFT }
\label{sec-CFT}

According to Kerr/CFT correspondence \cite{Guica:2008mu,extremeKerr-CFT,castro,hiddenkerrcft}, the four-dimensional physics of rotating black holes is holographic to two-dimensional CFT. The Green's functions for field perturbations with different spins have been proposed in \cite{Hartman:2009nz,Chen:2010ni}. In \cite{Becker:2012vda}, the authors found that the spin-$1/2$ Green's function can be obtained from the field theoretical technique by varying the boundary action with respect to the spinor fields. They also found that the field theoretical result is in agreement with what is expected from CFT calculation. The correlation function for the spinor operators in CFT is widely known from AdS/CFT correspondence \cite{mueck1}. 
The correlation function of conformal vector fields in Lorentz gauge has been obtained in the context of AdS$_{d+1}$/CFT$_{d}$ correspondence in \cite{mueck1,Freedman:1998tz} and in covariant gauge in \cite{Girotti:1999if}. One crucial point is that  the correlator vanishes for $d=2$ in Lorentz gauge \cite{mueck1,Freedman:1998tz,Girotti:1999if}.
 However we expect that the correlator of conformal vector operators must not vanish in Chandrasekhar gauge (\ref{chandra-gauge}). The reason is that we know the semiclassical absorption cross section of spin-1 fields in Kerr background is not zero  \cite{Hartman:2009nz,Chen:2010ni}. 
Moreover the correlation functions definitely depend on the gauge condition \cite{ElShowk:2011gz}. In fact, the general form for the correlator of conformal vector operators ${\cal O}_i$ (with conformal weight $\Delta$) read as
\be
\left\langle {{\cal O}_i(x){\cal O}_j(y) } \right\rangle  = \frac{{\mathcal{C}}}
{{\left| {x - y} \right|^{2\Delta } }}\left( {\eta_{ij}  + f_{ij} \left( {x,y} \right)} \right)
\label{OOCFT},
\ee
where $\mathcal{C}$ is a constant that depends on the number of dimensions of spacetime
and the functions $f_{ij}$ depend on the gauge condition.
Although the explicit form of functions $f_{ij}$ is known in Lorentz gauge \cite{mueck1} or covariant gauge \cite{Girotti:1999if}, however it is very unlikely to find $f_{ij}$'s in Chandrasekhar gauge (\ref{chandra-gauge}) due to the complicated structure of (\ref{chandra-gauge}). 
Nevertheless, inspired by the fact that the two-point function (\ref{2pt}) factorizes as $G_R{\cal Z}^{ij}$ to two factors ($G_R$ which is not sensitive to vector indices and ${\cal Z}^{ij}$ which depends on vector indices), we may associate the former factor to $\frac{{\mathcal{C}}}
{{\left| {x - y} \right|^{2\Delta } }}$ and the latter to ${\eta _{ij}  + f_{ij} \left( {x,y} \right)}$. In this regard, we consider the finite temperature correlation function on a torus with circumferences $1/T_L$ and $1/T_R$ \cite{Becker:2012vda} 
\be
\left\langle {\mathcal{O}{\mathcal{O}}} \right\rangle  \sim \left( {\frac{{\pi T_R }}
{{\sinh \left( {\pi T_R t_{12}^+ } \right)}}} \right)^{2h_R } \left( {\frac{{\pi T_L }}
{{\sinh \left( {\pi T_L t_{12}^- } \right)}}} \right)^{2h_L } \label{OOtorus}.
\ee
We note that  one can obtain the two-point function of scalars \cite{Hartman:2009nz,Chen:2010ni} and spin-$1/2$ fermions \cite{Becker:2012vda} just by plugging the suitable left and right conformal weights $h_R =h_L=\beta + 1/2$ and $h_R=\beta+1/2, h_L =\beta$ in (\ref{OOtorus}), respectively.

Analytic continuing $t$ to $it$ and assuming the integer frequencies $\omega  = 2\pi kT$ \cite{Becker:2012vda}, the Fourier transform of two-point function (\ref{OOtorus}) becomes 
\be
\widetilde{\left\langle {\mathcal{O}{\mathcal{O}}} \right\rangle}\sim {T_R}^{2\beta} \frac{{\Gamma \left( {1 - 2h_R } \right)\Gamma \left( {1 - 2h_L } \right)}}{{\Gamma \left( {1 - h_R  + {\textstyle{\frac{\omega_R}{2\pi T_R}}}} \right)\Gamma \left( {1 - h_R  - {\textstyle{\frac{\omega_R}{2\pi T_R}}}} \right)\Gamma \left( {1 - h_L  + {\textstyle{\frac{\omega_L}{2\pi T_L}}}} \right)\Gamma \left( {1 - h_R  - {\textstyle{\frac{\omega_L}{2\pi T_L}}}} \right)}}\label{JJ}
\ee
Identifying the frequencies as
\be
\frac{\omega_R}{2\pi T_R}  =  - in_R, ~\frac{\omega_L}{2\pi T_L}   =  - in_L,\label{idenCFT}
\ee
and the conformal weights as
\be
~~ h_R = \beta + 1/2,~h_L = \beta - 1/2,
\ee
and plugging into (\ref{JJ}) yields the two-point function 
\be
\widetilde{\left\langle {\mathcal{O}{\mathcal{O}}} \right\rangle}\sim T_R^{2\beta } \frac{{
\Gamma \left( { - 2\beta } \right)\Gamma \left( {\beta  + \tfrac{1}
{2} - in_R } \right)\Gamma \left( {\beta  - \tfrac{1}
{2} - in_L } \right)
}}
{{\sin \left( {2\pi \beta } \right)\Gamma \left( {2\beta } \right)\Gamma \left( {\tfrac{3}
{2} - \beta  - in_L } \right)\Gamma \left( {\tfrac{1}
{2} - \beta  - in_R } \right)}}\label{JJ1},
\ee
which is in agreement with (\ref{GreenF}) that was obtained by using the variational method. We note that to get (\ref{JJ1}), we absorb some terms of (\ref{JJ})  in the other part of two-point function that is associated to  ${\eta _{ij}  + f_{ij} \left( {x,y} \right)}$. 

\section{Concluding remarks}
\label{sec-conclu}

In this article, we have obtained the two-point function for the vector fields on the near horizon of near extremal Kerr black holes by varying the appropriate boundary action for the vector fields with respect to the boundary vector fields. 
One interesting result that emerges from the explicit calculation of the boundary action is that the degrees of freedom of boundary vector fields (which is two) supports the original idea of Kerr/CFT correspondence that the dual theory to the four-dimensional Kerr black hole is a two-dimensional CFT. 
This is in contrast to the well known AdS$_{d+1}$/CFT$_d$ result that the dimension of bulk theory is exactly one more than the dimension of dual CFT.
Moreover the two-point function for the vector fields factorizes in two terms. The first term is not sensitive to the vector indices while the second term depends on vector indices as well as the gauge condition. The structure of the two-point function is exactly in agreement with the correlator of vector operators in a CFT.
In deriving the two-point function of the vector fields, we have used some approximations and considered the leading terms of the boundary action. It is interesting to investigate the subleading terms of the boundary action to find their contributions to the two-point function and to their dual quantities in CFT. Moreover deriving the correlator of conformal vector operators in Chandrasekhar gauge is other interesting task. The dependence of Kerr/CFT correspondence on the gauge condition is the other open question. We address these issues in future article.  

\bigskip

{\Large Acknowledgments}\newline

This work was supported by the Natural Sciences and Engineering Research Council of Canada. HMS is supported by the Dean's scholarship from the college of graduate studies and research, University of Saskatchewan.\newline

\appendix
\section{Identities between operators ${\cal D}_n,{\cal D}_n^\dagger,{\cal L}_n,{\cal L}_n^\dagger$}\label{ap1}

A lengthy but straightforward calculation shows the following identities hold 

\begin{eqnarray}
\Delta \left( {\mathcal{D}_1  + \frac{1}{{\bar \rho ^ *  }}} \right)\left( {\mathcal{D}^\dag  _1  - \frac{1}{{\bar \rho ^ *  }}} \right) &=& \Delta \mathcal{D}_1 \mathcal{D}^\dag  _1  - \frac{{2iK}}{{\bar \rho ^ *  }} \label{ID1},
\\
\left( {\mathcal{L}^\dag  _0  + \frac{{ia\sin \theta }}{{\bar \rho ^ *  }}} \right)\left( {\mathcal{L}_1  - \frac{{ia\sin \theta }}{{\bar \rho ^ *  }}} \right)&=&  \mathcal{L}^\dag  _0 \mathcal{L}_1  + \frac{{2iQa\sin \theta }}{{\bar \rho ^ *  }}, \label{ID2}
\end{eqnarray}
and
\begin{eqnarray}
\Delta \left( {\mathcal{D}^\dag  _0  + \frac{1}{{\bar \rho ^ *  }}} \right)\left( {\mathcal{D}_0  - \frac{1}{{\bar \rho ^ *  }}} \right)&=& \Delta \mathcal{D}^\dag  _0 \mathcal{D}_0  + \frac{{2iK}}{{\bar \rho ^ *  }}, \label{ID3}
\\
\left( {\mathcal{L}_0  + \frac{{ia\sin \theta }}{{\bar \rho ^ *  }}} \right)\left( {\mathcal{L}^\dag  _1  - \frac{{ia\sin \theta }}{{\bar \rho ^ *  }}} \right) &=& \mathcal{L}_0 \mathcal{L}^\dag  _1  - \frac{{2iQa\sin \theta }}{{\bar \rho ^ *  }}.\label{ID4}
\end{eqnarray} 
where $K$ and $Q$ are given by (\ref{Kdef}) and (\ref{Qdef}) respectively.
\section{Chandrasekhar's solutions for Maxwell fields in Kerr spacetimes}\label{Apfullsol}
The full solutions for Maxwell fields in Kerr spacetime that include the gauge functions $H_\pm$, are given by \cite{Chandrasekhar:1985kt}, 

\begin{eqnarray}
A_t  &=& \frac{{ia}}{{\sqrt{2} \rho ^2 }}\left( {\left( {\Delta R_ +  \zeta _ +   - R_ -  \zeta _ -  } \right) - \left( {\xi _ +  S_ +   - \xi _ -  S_ -  } \right)\sin \theta } \right) \nonumber\\ 
  &+& \frac{1}{{\sqrt{2} \rho ^2}}\left( {\Delta \left( {{\cal D}_0 H_ +   - {\cal D}_0^\dag  H_ -  } \right) + ia\left( {{\cal L}_0^\dag  H_ +   - {\cal L}_0 H_ -  } \right)\sin \theta } \right), 
 \end{eqnarray}

\be
A_r  = \frac{{ia}}{{\sqrt 2 \Delta }}\left( {\Delta R_ +  \zeta _ +   + R_ -  \zeta _ -  } \right) + \frac{\Delta }{{\sqrt 2 }}\left( {{\cal D}_0 H_ +   + {\cal D}_0^\dag  H_ -  } \right),
\ee
\be
A_\theta   =  - \frac{1}{{\sqrt 2 }}\left( {\xi _ +  S_ +   + \xi _ -  S_ -  } \right) + \frac{1}{{\sqrt 2 }}\left( {{\cal L}_0^\dag  H_ +   + {\cal L}_0 H_ -  } \right),
\ee
and

\begin{eqnarray}
 A_\phi   &=&  - \frac{i}{{\sqrt 2 \rho ^2 }}\left( {a^2 \left( {\Delta R_ +  \zeta _ +   - R_ -  \zeta _ -  } \right)\sin ^2 \theta  - \left( {r^2  + a^2 } \right)\left( {\xi _ +  S_ +   - \xi _ -  S_ -  } \right)\sin \theta } \right) \nonumber\\ 
  &-& \frac{1}{{\sqrt 2 }}\left( {a\Delta \left( {{\cal D}_0 H_ +   - {\cal D}_0^\dag  H_ -  } \right)\sin ^2 \theta  + i\left( {r^2  + a^2 } \right)\left( {{\cal L}_0^\dag  H_ +   - {\cal L}_0 H_ -  } \right)\sin \theta } \right). 
 \end{eqnarray}

\section{Teukolsky equations and solutions in near region}\label{ap2}
The $R_\pm(r)$ and $S_ \pm(\theta)$ in Teukolsky wave function (\ref{PsiTeu}) satisfy the equations,
\be
\frac{{\partial _r \left( {\Delta ^{ \pm 1 + 1} \partial _r R_ \pm  } \right)}}
{{\Delta ^{ \pm 1} }}\left( {\frac{{H^2  \mp 2i\left( {r - M} \right)H}}
{\Delta } \pm 4i\omega r + 2am\omega  + 1 \pm 1 - K_l } \right)R_ \pm   = 0\label{Teu-rad},
\ee
which is known as the radial Teukolsky equation, and
\be
\frac{1}
{{\sin \theta }}\partial _\theta  \left( {\sin \theta \partial _\theta  S_ \pm \left( \theta  \right)} \right) - \left( {\frac{{m\left( {m \pm 2\cos \theta } \right) + 1}}
{{\sin ^2 \theta }} + a^2 \omega ^2 \sin ^2 \theta  \pm 2a\omega \cos \theta  - K_l } \right)S_ \pm \left( \theta  \right) = 0\label{Teu-ang},
\ee
which is the corresponding angular one for spin $\pm 1$. $K_l$ is the separation constant and $H = \omega (r^2+a^2) - am$. Following \cite{Hartman:2009nz}, for $x = (r-r_+)/r_+$, the radial equation for spin $\pm 1$ can be written as
\be
x\left( {x + \tau _H } \right)\partial _r \left( {\partial _r R_ \pm  } \right) + \left( {1 \pm 1} \right)\left( {2x + \tau _H } \right)\partial _r R_ \pm   + V_ \pm  R_ \pm   = 0\label{Teu-inx},
\ee
where
\[
V_ \pm   = \frac{{\left( {r_ +  \omega x^2  + 2r_ +  \omega x + \tfrac{1}
{2}n\tau _H } \right)^2  \mp i\left( {2x + \tau _H } \right)\left( {r_ +  \omega x^2  + 2r_ +  \omega x + \tfrac{1}
{2}n\tau _H } \right)}}
{{x\left( {x + \tau _H } \right)}}
\]
\be
 \pm 4ir_ +  \omega \left( {1 + x} \right) + 2am\omega  + 1 \pm 1 - K_l \label{Vteu-inx}.
\ee
The solution for ``near'' region is given in \cite{Hartman:2009nz} as
\be
R_ \pm ^{near}  = \left( {\frac{x}
{{\tau _H }} + 1} \right)^{i\left( {\tfrac{n}
{2} - m} \right) \mp 1} x^{ - \tfrac{{in}}
{2} \mp 1} {}_2F_1 \left( {\frac{1}
{2} + \beta  \mp 1 - im,\frac{1}
{2} - \beta  \mp 1 - im,1 \mp 1 - in, - \frac{x}
{{\tau _H }}} \right)\label{nearRsol},
\ee
where $\beta$ is given in (\ref{beta}). 
Considering only real and positive valued $\beta$ in (\ref{beta}) 
plays a role in deriving the corresponding two-point function from the variation of the boundary action (\ref{varSol}).

\section{Calculation of  ${\bf{A^\dagger \Xi A}}$ and the dominant terms}\label{ap3}

In this appendix we calculate explicitly all the terms of 
\begin{eqnarray}
{\bf{A^\dag\Xi A}}   &= &g^{rr} \frac{{\left( {{\bf{A}}_ + ^B } \right)^\dag  }}{{\left( {R_ + ^B } \right)^* }}R_ + ^* \left( {R_ +  ({\bf{\Pi }}_1 +{\bf{\Theta}}) } \right)\frac{{{\bf{A}}_ + ^B }}{{R_ + ^B }}
\nonumber\\
&+& g^{rr} \frac{{\left( {{\bf{A}}_ + ^B } \right)^\dag  }}{{\left( {R_ + ^B } \right)^* }}R_ + ^* \left( {R_ -  ({\bf{\Pi }}_3 +{\bf{\Theta}})  } \right){\bf{L\chi K}}^{ - 1} \frac{{{\bf{A}}_ + ^B }}{{R_ + ^B }} \nonumber\\
  &+& g^{rr} \frac{{\left( {{\bf{A}}_ + ^B } \right)^\dag  }}{{\left( {R_ + ^B } \right)^* }}\left( {R_ -  {\bf{L\chi K}}^{ - 1} } \right)^\dag  \left( {R_ +  ({\bf{\Pi }}_1 +{\bf{\Theta}}) } \right)\frac{{{\bf{A}}_ + ^B }}{{R_ + ^B }}
\nonumber\\
  &+& g^{rr} \frac{{\left( {{\bf{A}}_ + ^B } \right)^\dag  }}{{\left( {R_ + ^B } \right)^* }}\left( {R_ -  {\bf{L\chi K}}^{ - 1} } \right)^\dag  \left( {R_ -  ({\bf{\Pi }}_3 +{\bf{\Theta}})} \right){\bf{L\chi K}}^{ - 1} \frac{{{\bf{A}}_ + ^B }}{{R_ + ^B }} \label{AAcalnoQ},
\end{eqnarray}
that is the result of equation (\ref{ADAcal}) after dropping the terms proportional to $Q_\pm$. The different components of $\bf{A}_+^B$ are

\begin{eqnarray}
 A_{t + }^B  &=& e^{-i\omega t + im\phi}\left( {f_1 S_ -   + f_2 S_ +  } \right)R_ +^B  \Delta_B, \nonumber \\ 
 A_{r + }^B  &=& e^{-i\omega t + im\phi}\left( {f_5 S_ -   + f_6 S_ +  } \right)R_ +^B, \nonumber \\ 
 A_{\theta  + }^B  &=& e^{-i\omega t + im\phi}\left( {f_9 S_ -   + f_{10} S_ +  } \right)R_ +^B  \Delta_B, \nonumber\\ 
 A_{\phi  + }^B  &=& e^{-i\omega t + im\phi}\left( {f_{13} S_ -   + f_{14} S_ +  } \right)R_ +^B  \Delta _B. \label{AAapd}
 \end{eqnarray}

Each term of (\ref{AAcalnoQ}) can be rewritten as
\par
\be
g^{rr} \frac{{\left( {{\bf{A}}_ + ^B } \right)^\dag  }}{{\left( {R_ + ^B } \right)^* }}R_ + ^* \left( {R_ + {\bf{\Pi }}_1 } \right)\frac{{{\bf{A}}_ + ^B }}{{R_ + ^B }} = g^{rr} \frac{{R_ + ^* R_ +  }}{{\left( {R_ + ^B } \right)^* R_ + ^B }}\pi _1^{ij} A_{i+}^{B*} A_{j+}^B ,
\ee
\be
g^{rr} \frac{{\left( {{\bf{A}}_ + ^B } \right)^\dag  }}{{\left( {R_ + ^B } \right)^* }}R_ + ^* \left( {R_ + {\bf{\Theta }} } \right)\frac{{{\bf{A}}_ + ^B }}{{R_ + ^B }} = g^{rr} \frac{{R_ + ^* R_ +  }}{{\left( {R_ + ^B } \right)^* R_ + ^B }}\theta _1^{ij} A_{i+}^{B*} A_{j+}^B,
\ee
\be
g^{rr} \frac{{\left( {{\bf{A}}_ + ^B } \right)^\dag  }}{{\left( {R_ + ^B } \right)^* }}R_ + ^* \left( {R_ -  {\bf{\Pi }}_3  } \right){\bf{L\chi K}}^{ - 1} \frac{{{\bf{A}}_ + ^B }}{{R_ + ^B }} = g^{rr} \frac{{R_ + ^* R_ -  }}{{\left( {R_ + ^B } \right)^* R_ + ^B }}\pi _2^{ij} A_{i+}^{B*} A_{j+}^B,
\ee

\be
g^{rr} \frac{{\left( {{\bf{A}}_ + ^B } \right)^\dag  }}{{\left( {R_ + ^B } \right)^* }}R_ + ^* \left( {R_ -  {\bf{\Theta}}  } \right){\bf{L\chi K}}^{ - 1} \frac{{{\bf{A}}_ + ^B }}{{R_ + ^B }} = g^{rr} \frac{{R_ + ^* R_ -  }}{{\left( {R_ + ^B } \right)^* R_ + ^B }}\theta _2^{ij} A_{i+}^{B*} A_{j+}^B,
\ee

\be
g^{rr} \frac{{\left( {{\bf{A}}_ + ^B } \right)^\dag  }}{{\left( {R_ + ^B } \right)^* }}\left( {R_ -  {\bf{L\chi K}}^{ - 1} } \right)^\dag  \left( {R_ +  {\bf{\Pi }}_1 } \right)\frac{{{\bf{A}}_ + ^B }}{{R_ + ^B }} = g^{rr} \frac{{R_ - ^* R_ +  }}{{\left( {R_ + ^B } \right)^* R_ + ^B }}\pi _3^{ij} A_{i+}^{B*} A_{j+}^B
,\ee

\be
g^{rr} \frac{{\left( {{\bf{A}}_ + ^B } \right)^\dag  }}{{\left( {R_ + ^B } \right)^* }}\left( {R_ -  {\bf{L\chi K}}^{ - 1} } \right)^\dag  \left( {R_ + {\bf{\Theta}} } \right)\frac{{{\bf{A}}_ + ^B }}{{R_ + ^B }} = g^{rr} \frac{{R_ - ^* R_ +  }}{{\left( {R_ + ^B } \right)^* R_ + ^B }}\theta _3^{ij} A_{i+}^{B*} A_{j+}^B 
,\ee

\be
g^{rr} \frac{{\left( {{\bf{A}}_ + ^B } \right)^\dag  }}{{\left( {R_ + ^B } \right)^* }}\left( {R_ -  {\bf{L\chi K}}^{ - 1} } \right)^\dag  \left( {R_ -  {\bf{\Pi }}_3} \right){\bf{L\chi K}}^{ - 1} \frac{{{\bf{A}}_ + ^B }}{{R_ + ^B }} = g^{rr} \frac{{R_ - ^* R_ -  }}{{\left( {R_ + ^B } \right)^* R_ + ^B }}\pi _4^{ij} A_{i+}^{B*} A_{j+}^B
,\ee

\be
g^{rr} \frac{{\left( {{\bf{A}}_ + ^B } \right)^\dag  }}{{\left( {R_ + ^B } \right)^* }}\left( {R_ -  {\bf{L\chi K}}^{ - 1} } \right)^\dag  \left( {R_ - {\bf{\Theta}}} \right){\bf{L\chi K}}^{ - 1} \frac{{{\bf{A}}_ + ^B }}{{R_ + ^B }} = g^{rr} \frac{{R_ - ^* R_ -  }}{{\left( {R_ + ^B } \right)^* R_ + ^B }}\theta _4^{ij} A_{i+}^{B*} A_{j+}^B
,\ee
where
\be
\pi _1^{ij} A_{i+}^{B*} A_{j+}^B =\left( {g^{tt} A_{t +} ^{B*}  + g^{t\phi } A_{\phi +} ^{B*} } \right)A_t ^B  
+ \left( {g^{t\phi } A_{t +} ^{B*}  + g^{\phi \phi } A_{\phi +} ^{B*} } \right)A_\phi^B  \label{pi1},
\ee
\be
\theta _1^{ij} A_{i+}^{B*} A_{j+}^B = - \left( {\left( {-i\omega g^{\phi t}  + im g^{\phi \phi }  } \right)A_{\phi +} ^{B*}  
+ \left( {-i\omega g^{tt}  + im g^{t\phi }  } \right)A_{t +} ^{B*} } \right)A_r^B   \label{tt1}
,\ee
\begin{eqnarray}
\pi _2^{ij} A_{i+}^{B*} A_{j+}^B   &=& \frac{{ - 1}}{{\Delta_B \left( {f_{13} f_2  - f_1 f_{14} } \right)}}\left( {A_{t +} ^{B} A_{t +} ^{B*} \left( {g^{tt} \left( {f_4 f_{13}  - f_3 f_{14} } \right) + g^{t\phi } \left( {f_{16} f_{13}  - f_{15} f_{14} } \right)} \right)} \right. \nonumber\\ 
  &+& A_{t +} ^{B} A_{\phi +} ^{B*} \left( {g^{t\phi } \left( {f_4 f_{13}  - f_3 f_{14} } \right) + g^{\phi \phi } \left( {f_{16} f_{13}  - f_{15} f_{14} } \right)} \right)\nonumber \\ 
  &+& A_{\phi +} ^{B}  A_{t +} ^{B*} \left( {g^{tt} \left( {f_3 f_2  - f_4 f_1 } \right) + g^{t\phi } \left( {f_{15} f_2  - f_{16} f_1 } \right)} \right) \nonumber\\ 
 &+&\left. {  A_{\phi +} ^{B}  A_{\phi +} ^{B*} \left( {g^{t\phi } \left( {f_3 f_2  - f_4 f_1 } \right) + g^{\phi \phi } \left( {f_{15} f_2  - f_{16} f_1 } \right)} \right)} \right), \label{pi2}
 \end{eqnarray}
\begin{eqnarray}
\theta _2^{ij} A_{i+}^{B*} A_{j+}^B &=& \frac{{\left( {A_{t+}^B \left( {f_8 f_{13}  - f_7 f_{14} } \right) + A_{\phi+}^B  \left( {f_7 f_2  - f_8 f_1 } \right)} \right)}}{{\Delta_B^2 \left( {f_1 f_{14}  - f_{13} f_2 } \right)}}\nonumber\\
& \times&
\left( {A_{t +} ^{B*} \left( { - i\omega g^{tt}  + img^{t\phi } } \right) + A_{\phi +} ^{B*} \left( { - i\omega g^{t\phi }  + img^{\phi \phi } } \right)} \right) \label{tt2}
,\end{eqnarray}

\begin{eqnarray}
\pi _3^{ij} A_{i+}^{B*} A_{j+}^B   &=& \frac{{ - (\beta -3/2)}}{{\Delta_B r_B\left( {f_1^* f_{14}^*  - f_{13} f_2 } \right)}}\left( {A_{t +} ^{B} A_{t +} ^{B*} \left( {g^{tt} \left( {f_3^* f_{14}^*  - f_4^* f_{13}^* } \right) + g^{t\phi } \left( {f_5^* f_{14}^*  - f_6^* f_{13}^* } \right)} \right)} \right. \nonumber\\ 
  &+& A_{t +} ^{B} A_{\phi +} ^{B*} \left( {g^{tt} \left( {f_4^* f_1^*  - f_3^* f_2^* } \right) + g^{t\phi } \left( {f_{16}^* f_1^*  - f_{15}^* f_2^* } \right)} \right)\nonumber \\ 
  &+& A_{\phi +} ^{B} A_{t +} ^{B*} \left( {g^{t\phi } \left( {f_3^* f_{14}^*  - f_4^* f_{13}^* } \right) + g^{t\phi } \left( {f_{15}^* f_{14}^*  - f_{16}^* f_{13}^* } \right)} \right) \nonumber\\ 
 &+&\left. {  A_{\phi +} ^{B} A_{\phi +} ^{B*} \left( {g^{tt} \left( {f_4^* f_1^*  - f_3^* f_2^* } \right) + g^{t\phi } \left( {f_{16}^* f_1^*  - f_{15}^* f_2^* } \right)} \right)} \right), \label{pi3}
 \end{eqnarray} 

\begin{eqnarray}
\theta _3^{ij} A_{i+}^{B*} A_{j+}^B  &=& \frac{{ - 1}}{{\Delta_B \left( {f_1^* f_{14}^*  - f_{13} f_2 } \right)}}\left( {A_{r +} ^{B} A_{t +} ^{B*} \left( { - i\omega g^{tt} \left( {f_4^* f_{13}^*  - f_3^* f_{14}^* } \right) + img^{\phi \phi } \left( {f_{16}^* f_3^*  - f_{15}^* f_{14}^* } \right)} \right.} \right. \nonumber\\ 
&+& \left. { g^{t\phi } \left( { - i\omega \left( {f_{16}^* f_3^*  - f_{15}^* f_{14}^* } \right) + im\left( {f_4^* f_{13}^*  - f_3^* f_{14}^* } \right)} \right)} \right) \nonumber\\ 
  &+& A_{r +} ^{B} A_{\phi +} ^{B*} \left( { - i\omega g^{tt} \left( {f_3^* f_2^*  - f_4^* f_1^* } \right) + img^{\phi \phi } \left( {f_{15}^* f_2^*  - f_6^* f_1^* } \right)} \right.\nonumber \\ 
 &+&\left. {  g^{t\phi } \left( { - i\omega \left( {f_{15}^* f_2^*  - f_{16}^* f_1^* } \right) + im\left( {f_3^* f_2^*  - f_4^* f_1^* } \right)} \right)} \right) \label{tt3},
 \end{eqnarray} 

\begin{eqnarray}
  \pi _4^{ij} A_{i+}^{B*} A_{j+}^B &=& \frac{{ - (\beta + 1/2)}}{{\Delta_B ^2 r_B\left| {f_1 f_{14}  - f_{13} f_2 } \right|^2 }}\left( {A_{\phi +} ^{B} A_{\phi +} ^{B*} \left( {g^{tt} \left( {f_3^* f_2^* f_4 f_1  - f_3^* f_2^* f_3 f_2  - f_4^* f_1^* f_4 f_1  + f_4^* f_1^* f_3 f_2 } \right)} \right.} \right. \nonumber\\ 
  &+& g^{\phi \phi } \left( {f_{15}^* f_2^* f_{16} f_1  + f_{16}^* f_1^* f_{15} f_2  - f_{15}^* f_2^* f_{15} f_2  - f_{16}^* f_1^* f_{16} f_1 } \right) \nonumber\\ 
  &+& g^{t\phi } \left( {f_{16}^* f_1^* f_3 f_2  - f_4^* f_1^* f_{16} f_1  + f_3^* f_2^* f_{16} f_1  - f_{16}^* f_1^* f_4 f_1 } \right. \nonumber\\ 
 &-&\left. { f_3^* f_2^* f_{15} f_2  - f_{15}^* f_2^* f_3 f_2  + f_4^* f_1^* f_{15} f_2  + f_{15}^* f_2^* f_4 f_1 } \right) \nonumber\\ 
  &+& A_{\phi +} ^{B} A_{t +} ^{B*} \left( {g^{tt} \left( {f_4^* f_{13}^* f_4 f_1  + f_3^* f_{14}^* f_3 f_2  - f_3^* f_{14}^* f_4 f_1  - f_4^* f_{13}^* f_3 f_2 } \right)} \right. \nonumber\\ 
  &+& g^{\phi \phi } \left( {f_{15}^* f_{14}^* f_{15} f_2  + f_{16}^* f_{13}^* f_{16} f_1  - f_{15}^* f_{14}^* f_{16} f_1  - f_{16}^* f_{13}^* f_{15} f_2 } \right) \nonumber\\ 
  &+& g^{t\phi } \left( {f_{15}^* f_{14}^* f_3 f_2  - f_{15}^* f_{14}^* f_4 f_1  - f_4^* f_{13}^* f_{15} f_2  + f_{16}^* f_{13}^* f_4 f_1 } \right. \nonumber\\
 &+&\left. {  f_3^* f_{14}^* f_{15} f_2  - f_{16}^* f_{13}^* f_3 f_2  - f_3^* f_{14}^* f_{16} f_1  + f_4^* f_{13}^* f_{16} f_1 } \right) \nonumber\\ 
  &+& A_{t +} ^{B} A_{t +} ^{B*} \left( {g^{tt} \left( {f_3^* f_{14}^* f_4 f_{13}  + f_4^* f_{13}^* f_3 f_{14}  - f_3^* f_{14}^* f_3 f_{14}  - f_4^* f_{13}^* f_4 f_{13} } \right)} \right. \nonumber\\ 
  &+& g^{\phi \phi } \left( {f_{16}^* f_{13}^* f_{15} f_{14}  + f_{15}^* f_{14}^* f_{16} f_{13}  - f_{16}^* f_{13}^* f_{16} f_{13}  - f_{15}^* f_{14}^* f_{15} f_{14} } \right) \nonumber\\ 
  &+& g^{t\phi } \left( {f_{15}^* f_{14}^* f_4 f_{13}  - f_3^* f_{14}^* f_{15} f_{14}  - f_{15}^* f_{14}^* f_3 f_{14}  + f_4^* f_{13}^* f_{15} f_{14} } \right. \nonumber\\ 
 &+&\left. {  f_{16}^* f_{13}^* f_3 f_{14}  - f_{16}^* f_{13}^* f_4 f_{13}  + f_3^* f_{14}^* f_{16} f_{13}  - f_4^* f_{13}^* f_{16} f_{13} } \right) \nonumber\\ 
  &+& A_{t +} ^{B} A_{\phi +} ^{B*} \left( {g^{tt} \left( {f_3^* f_2^* f_3 f_{14}  + f_4^* f_1^* f_4 f_{13}  - f_4^* f_1^* f_3 f_{14}  - f_3^* f_2^* f_4 f_{13} } \right)} \right. \nonumber\\ 
  &+& g^{\phi \phi } \left( {f_{15}^* f_2^* f_{15} f_{14}  - f_{15}^* f_2^* f_{16} f_{13}  + f_{16}^* f_1^* f_{16} f_{13}  - f_{16}^* f_1^* f_{15} f_{14} } \right) \nonumber\\ 
 &+&g^{t\phi } \left( {f_{15}^* f_2^* f_3 f_{14}  - f_{16}^* f_1^* f_3 f_{14}  - f_{15}^* f_2^* f_4 f_{13}  - f_4^* f_1^* f_{15} f_{14} } \right. \nonumber\\ 
 &+&\left.{\left. {\left. {  f_4^* f_1^* f_{16} f_{13}  + f_{16}^* f_1^* f_4 f_{13}  - f_3^* f_2^* f_{16} f_{13}  + f_3^* f_2^* f_{15} f_{14} } \right)} \right)}\right),\label{pi4}
 \end{eqnarray}

\begin{eqnarray}
\theta _4^{ij} A_{i+}^{B*} A_{j+}^B &=& \frac{1}{{\Delta_B ^3\left| {f_1 f_{14}  - f_{13} f_2} \right|^2 }}\left( {A_{t + }^{B*} A_{t + }^B \left( {f_7 f_{14}  - f_8 f_{13} } \right)\left( {im\left( {\left( {f_{16}^* f_{13}^*  - f_{15}^* f_{14}^* } \right)g^{\phi \phi }  + \left( {f_4^* f_{13}^*  - f_3^* f_{14}^* } \right)g^{t\phi } } \right)} \right.} \right. \nonumber\\ 
 &-&\left. { i\omega \left( {\left( {f_{16}^* f_{13}^*  - f_{15}^* f_{14}^* } \right)g^{\phi t}  - \left( {f_4^* f_{13}^*  - f_3^* f_{14}^* } \right)g^{tt} } \right)} \right) + A_{t + }^{B*} A_{\phi  + }^B \left( {f_7 f_{14}  - f_8 f_{13} } \right) \nonumber\\ 
  &\times& \left( {im\left( {\left( {f_{15}^* f_2^*  - f_{16}^* f_1^* } \right)g^{\phi \phi }  + \left( {f_3^* f_2^*  - f_4^* f_1^* } \right)g^{t\phi } } \right) - i\omega \left( {\left( {f_{15}^* f_2^*  - f_{16}^* f_1^* } \right)g^{\phi t}  - \left( {f_4^* f_1^*  - f_3^* f_2^* } \right)g^{tt} } \right)} \right)\nonumber \\ 
  &+& A_{\phi  + }^{B*} A_{t + }^B \left( {f_8 f_1  - f_2 f_7 } \right)\left( {im\left( {\left( {f_{16}^* f_{13}^*  - f_{15}^* f_{14}^* } \right)g^{\phi \phi }  + \left( {f_4^* f_{13}^*  - f_3^* f_{14}^* } \right)g^{t\phi } } \right)} \right. \nonumber\\ 
 &-&\left. {  i\omega \left( {\left( {f_{16}^* f_{13}^*  - f_{15}^* f_{14}^* } \right)g^{\phi t}  - \left( {f_4^* f_{13}^*  - f_3^* f_{14}^* } \right)g^{tt} } \right)} \right) + A_{\phi  + }^{B*} A_{\phi  + }^B  \nonumber\\ 
  &\times& \left( {f_8 f_1  - f_2 f_7 } \right)\left( {im\left( {\left( {f_{15}^* f_2^*  - f_{16}^* f_1^* } \right)g^{\phi \phi }  + \left( {f_3^* f_2^*  - f_4^* f_1^* } \right)g^{t\phi } } \right)} \right.\nonumber \\ 
 &-&\left. {\left. {  i\omega \left( {\left( {f_{15}^* f_2^*  - f_{16}^* f_1^* } \right)g^{\phi t}  - \left( {f_4^* f_1^*  - f_3^* f_2^* } \right)g^{tt} } \right)} \right)} \right) \label{tt4}.
 \end{eqnarray}

To obtain (\ref{pi1}) - (\ref{tt4}), we consider only the terms that couple to $g^{tt}$, $g^{t\phi}$ and  $g^{\phi\phi}$ because they are the leading order terms compared to the terms that couple to $g^{rr}$ or $g^{\theta\theta}$. This fact can be seen from equations (\ref{contra-metric1}) and  (\ref{contra-metric2}) where $\Delta=\Delta_B$ is a very small number. 
A simple analysis of eight equations (\ref{pi1}) - (\ref{tt4}) shows that $\pi_4^{ij}$ and $\theta_4^{ij}$ are the dominant terms in (\ref{AAcalnoQ}). Both terms are in order of $\Delta_B^{-3}$ compared to $\pi_2$, $\theta_2$, $\pi_3$ and $\theta_3$ that are in order of  $\Delta_B^{-2}$ and $\pi_1$ and $\theta_1$ are in order of $\Delta_B^{-1}$. We show the summation of the dominant terms $\pi_4^{ij}$ and $\theta_4^{ij}$ by 
\begin{equation}
\tilde \theta_4^{ij}=\pi_4^{ij}+\theta_4^{ij}.\label{tildetheta}
\end{equation}

\end{document}